# Bootstrapping Life-Inspired Machine Intelligence: The Biological Route from Chemistry to Cognition and Creativity

*Giovanni Pezzulo,*[1,*] *and Michael Levin*[2,3,*]

1 Institute of Cognitive Sciences and Technologies, National Research Council, Rome, Italy
2 Allen Discovery Center at Tufts University, Medford, MA 02155, USA.
3 Wyss Institute for Biologically Inspired Engineering, Harvard University, Boston, MA 02115, USA.

Corresponding authors: giovanni.pezzulo@istc.cnr.it, michael.levin@tufts.edu

**Declarations:**

**Author Contributions:** G.P. and M.L. wrote the manuscript text and prepared the figures together. All authors reviewed the manuscript.

**Competing Interests:** M.L.'s lab receives a sponsored research agreement from Astonishing Labs, a company seeking to operate in the space of non-neuromorphic, bio-inspired AI.

## Abstract

Achieving advanced machine intelligence remains a central challenge in AI research, often approached through scaling neural architectures and generative models. However, biological systems offer a broader repertoire of strategies for adaptive, goal-directed behavior—strategies that emerged long before nervous systems evolved. This paper advocates a genuinely life-inspired approach to machine intelligence, drawing on principles from biology that enable robustness, autonomy, and open-ended problem-solving across scales. We frame intelligence as flexible problem-solving, following William James, and develop the concept of "cognitive light cones" to characterize the continuum of intelligence in living systems and machines. We argue that biological evolution has discovered a scalable recipe for intelligence—and the progressive expansion of organisms' "cognitive light cone", predictive and control capacities. To explain how this is possible, we distill five design principles— multiscale autonomy, growth through self-assemblage of active components, continuous reconstruction of capabilities, exploitation of physical and embodied constraints, and pervasive signaling enabling self-organization and top-down control from goals—that underpin life's ability to navigate creatively diverse problem spaces. We discuss how these principles contrast with current AI paradigms and outline pathways for integrating them into future autonomous, embodied, and resilient artificial systems.

## Introduction

We live in interesting times: a historical period marked by a widespread confidence that achieving machine intelligence may be within reach. This has sparked intense debate on a range of topics, some of which are primarily theoretical [1,2]. These include whether artificial systems are already endowed—or will soon be endowed—with forms of intelligence (or even general intelligence), autonomy, understanding, or consciousness (and what these terms actually mean); how such abilities compare to those found in biological organisms at different levels of complexity; and what the implications of these technological advances may be for science, society, and ethics.

Other questions are more technical in nature, such as which methods might be most effective for achieving advanced machine intelligence. Examples include scaling up the currently most successful technologies based on generative (possibly multimodal) AI; enabling autonomous exploration and learning through interaction with the environment to surpass human skill levels; supporting interaction with others to develop shared understanding, cooperative behaviors, and alignment; and to what extent taking inspiration from the neural mechanisms supporting cognition and intelligence in living organisms matters [2–4].

In parallel, an emerging body of research spanning not only cognitive science and neuroscience, but also theoretical biology, molecular biology, and genetics—while also intersecting with artificial intelligence and robotics—aims to understand intelligence across substrates, scales, and origin stories. This line of work, the field of Diverse Intelligence [5–9], investigates a wide diversity of agents, including biological organisms, artificial systems, robots, cyborgs, and forms of active matter, emphasizing the principles that underlie adaptive, goal-directed behavior across various manifestations and substrates. However, the field of machine intelligence [10] has, as a whole, not yet sufficiently benefitted from the insights of a broad approach to basal cognition in diverse substrates.

This perspective advances a genuinely life-inspired approach to machine intelligence. Rather than focusing on neural computation alone, it draws on the broader repertoire of problem-solving strategies evolved by living systems, revealing remarkable capabilities—such as self-organization, robustness, and open-ended adaptation—that emerge from the dynamics of life itself and were present long before nervous systems evolved [11,12].

This paper discusses how natural forms self-assemble—from chemistry to advanced cognitive systems—and distills design principles that may prove useful for the development of future life-inspired intelligent machines, autonomous and embodied AIs, and robots. We argue that biology has not only already "built" many forms of intelligence, including ourselves, but has also discovered a recipe for *flexible* and *scalable* intelligence. We discuss the key design principles that make this possible and the implications for machine intelligence.

In the remainder of this paper, we first provide a general introduction to the concept of diverse intelligence, illustrating examples from biology and connecting to machine intelligence. We then introduce five design principles drawn from biology. Finally, we discuss them in the context of current approaches in machine learning, (embodied) AI, and robotics, with the aim of identifying points of convergence, divergence, and promising future directions. We aim to answer questions like: why don't we have synthetic systems that achieve the same level of flexible and scalable problem-solving as living organisms? What are the knowledge gaps? How can we leverage nature's design principles to pursue more advanced and general forms of AI?

## Diverse intelligence in biology and machines

How can intelligence be defined in a way that is not anchored to our (human) case, but instead applies across the breadth of biological systems—and potentially to machines—without becoming so broad as to lose explanatory power? While intelligence is notoriously difficult and controversial to define, a useful starting point is a definition attributed to William James: "Intelligence is a fixed goal with variable means of achieving it."

According to this view, intelligence is characterized by the capacity to solve problems flexibly (i.e., in many ways, including ways for which they were not directly trained), generalizing to novel contexts ("out of distribution," in machine learning parlance). Human problem-solving abilities are typically flexible, even if bounded. For example, we are generally able to obtain water (the goal) in both familiar and unfamiliar situations, using a variety of means: by relying on well-known routines (e.g., opening a refrigerator, asking a waiter in a restaurant, buying a bottle at a supermarket, or drinking from a river), but also by planning and imagining novel or creative solutions (e.g., lighting a fire to attract the attention of a passing ship when stranded and thirsty on a deserted island).

However, humans are not the only "intelligent" systems. First, cognitive capacities are spread along a spectrum with many different competencies [13]. Second, biological systems at all scales have been performing the sensing-decision-making-actuation loop in many (nested) problem spaces [14] besides the familiar 3D space colonized by nerve and muscle (Figure 1). This becomes apparent when we consider where "humans" come from. Both on an individual scale, and on an evolutionary scale, we all start life as an oocyte – an egg cell which looks to be the subject of chemistry and physics. Embryology offers *no* discrete process or step during which a blob of chemicals suddenly becomes mind-ful. What evolution and developmental biology reveal instead is a continuous, gradual process in which a self-assembling material slowly traverses across chemistry, physiology, developmental biology, neuroscience, behavioral science, and psychology (and eventually psychotherapy). While we have given discrete names to disciplines (and journals and university departments), nature teaches the need for models of transformation and scaling – of projecting navigational competencies across problem spaces.

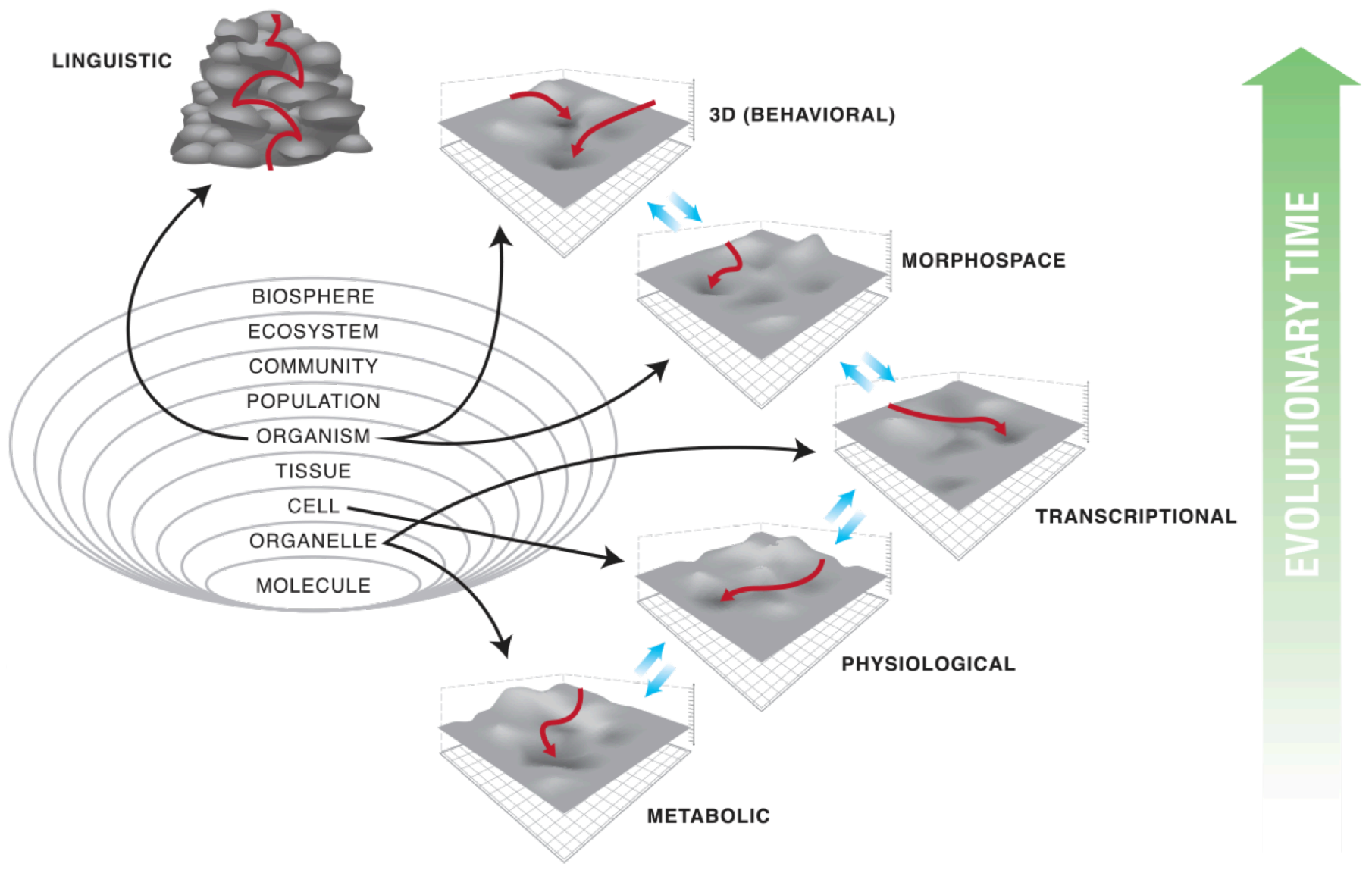

***Figure 1. Living beings, from simpler to more complex, navigate across nested problem spaces.*** *Evolution pivoted systems across a variety of problem spaces, starting with metabolic and physiological state spaces, followed by transcriptional space once genes came on the scene,*

*anatomical morphospace once multicellularity appeared, then conventional behavior in 3D space (once nerve and muscle evolved), and now exploring linguistic and many other problem spaces inhabited by humans with culture and intelligent artifacts. It is likely that many of the same policies and problem-solving skills were adapted to novel spaces by active agents that needed to navigate these spaces to meet adaptive goals in the face of novel stresses and opportunities. Panel taken with permission from* [14].

Framing intelligence in terms of flexible problem-solving highlights that intelligence is far more diverse than is often assumed, manifesting in multifarious and sometimes surprising forms throughout biology. Many living systems—not only complex animals like ourselves, but also simpler animals, individual cells, and even subcellular gene regulatory networks—can be understood as flexible problem-solving systems. The problems they solve are shaped by their specific ecological niches: simple organisms typically address relatively short-sighted homeostatic goals and have a limited range of solutions (although they are not as trivial as commonly assumed). More complex organisms instead address more complex problems and can imagine and achieve more complex goals. These problems lie in problem spaces that are unconventional from our (human) perspective; for example, a group of cells that self-organize to create a body morphology is solving a problem in navigating a "morphospace" that differs from the typical spaces that we humans face, e.g. from spatial navigation problems in 3D physical space.

Crucially, within the boundaries defined by their ecological niches, problem spaces and resources (e.g., energy, time, memory), living organisms show *flexible* problem solving. Key to survival is their capability to deploy their abilities in uncertain and changing conditions – which therefore cannot be fully specified a priori, e.g., at the level of the genetically-encoded hardware. Changes are ubiquitous in nature and intelligence is unavoidably linked to the capacity to continuously adapt to novel and unforeseen situations. For example, tadpoles of the frog Xenopus laevis can flexibly solve sensorimotor problems even when their sensory organs are radically reconfigured. Experiments have shown that when eyes are transplanted to ectopic locations—such as the tail—the eyes become functional even though not connected directly to the brain; these tadpoles are able to use the visual information provided by the misplaced eyes to guide light-mediated learning and behavior, without new rounds of mutation and selection to accommodate this novel architecture. This example (and others shown in Figure 2) illustrates how biological intelligence supports flexible problem-solving by reassigning function, reinterpreting signals, and sustaining goal-directed behavior even under profound changes in bodily structure.

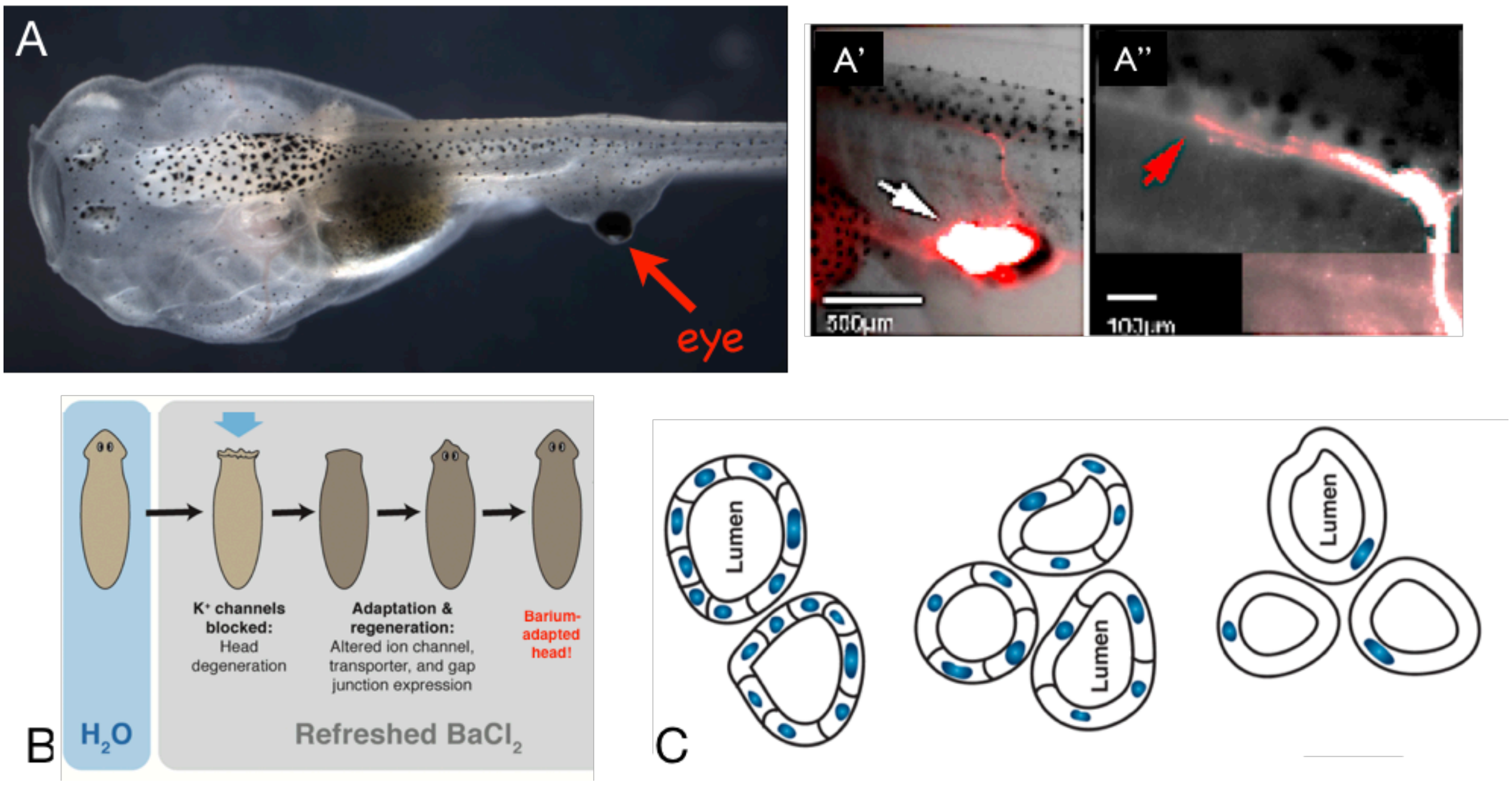

***Figure 2. Examples of diverse intelligence – understood as flexible and scalable problem-solving – in biology.*** *(A) Tadpoles of the frog Xenopus laevis can be produced with eyes on their tale (red arrowhead). These eyes (white arrowhead) produce optic nerves (red arrowheads) that connect to the spinal cord (A', and close-up in A") not the brain; nevertheless, visual learning assays indicate that these animals have functional vision* [15], *despite a novel sensory motor architecture and no new rounds of mutation or selection needed to make it work. (B) Planarian flatworms exposed to the nonspecific potassium channel blocker Barium experience rapid destruction of their heads, but then regenerate heads that manage to operate their neural and other cells' physiology despite the presence of barium* [16]*; these barium-adapted heads identified just a handful of new genes out of tens of thousands of possible moves in transcriptional space, which solve the physiological stressor they have not experienced in their history. (C) Cross-sections of kidney tubules in newts normally consist of 8-10 cells cooperating to produce the structure with a lumen; when cells are artificially created with multiple copies of their genome it is observed* [17,18] *that: cells adjust their size to their nucleus volume, a normal newt is produced despite extra copies of genetic material, and fewer cells cooperate to make the exact same structure, adjusting the size of the body to the novel cell size. In fact, when the cells are made truly enormous, just 1 cell will bend around itself, leaving a lumen in the middle, showing how different molecular mechanisms (cell:cell communication, or cytoskeletal bending) can be called up in order to solve an anatomical problem (create a species-specific target morphology when the system's components are unexpectedly changed) as needed. Panels B and C courtesy of Jeremy Guay of Peregrine Creative. Images used with permission from the references indicated.*

### *The "cognitive light cone" of living organisms and its continuous expansion*

Defining intelligence as flexible problem-solving reveals a continuum of problem-solving capacities in living organisms. One way to characterize this continuum is by appealing to the concept of a *cognitive light cone* [19], that is, a functional boundary of intelligence representing the scale and limits of an agent's goals, memory, predictive abilities, and problem-solving capacity across space and time (Figure 3). Specifically, the cognitive light cone is designed to represent the *size of the goals* that a given system can represent and pursue (e.g., in space and time). This captures the intuition that one can tell the level of sophistication of an agent by analyzing the scope of their goals. At the lowest level, the cognitive light cone indicates the size of the setpoints pursued by a homeostatic system, i.e., a “goal” is not necessarily a meta-cognitive “I know what my goals are” but simply the formalization of the cybernetic notion of navigating a space toward a preferred state [13].

Note that while the illustration of a bidirectional cognitive light cone spanning the two axes of time and space is a useful approximation, the cognitive light cone should be understood as a multidimensional construct. It encompasses not only the spatial range of control and temporal depth into the past (memory) and future (anticipation), but also additional dimensions, such as the size and complexity of pursued goals; the diversity of means (e.g., behavioral policies) available to achieve them; the structure and complexity of relevant problem spaces; and the range of possible outcomes or other entities productively included within the agent’s “radius of care” [20]. These dimensions are related but do not necessarily co-vary, and organisms may exhibit strong capabilities along some dimensions while remaining limited along others. For example, a simple animal such as a honey bee might have a sophisticated capacity to pursue navigational goal states spanning a relatively large spatial range while exhibiting limited temporal depth.

As a first approximation, the more extended an agent’s cognitive light cone, the greater its ability to flexibly solve increasingly challenging problems. For example, the cognitive light cone of a bacterium tracks local sugar concentration, while the cone of a human also allows them to not only store and buy sugar in order to prepare a cake, but to work toward stabilization of global food prices in the coming century (cognitive light cone that extends beyond the agent’s lifespan). Between those examples lie a massive spectrum of agents, such as for example amphibian cellular collectives whose anatomical setpoint goals (“build a limb, and rebuild it if it gets damaged) seem grandiose and incomprehensible to the cells (and the molecular networks within them).

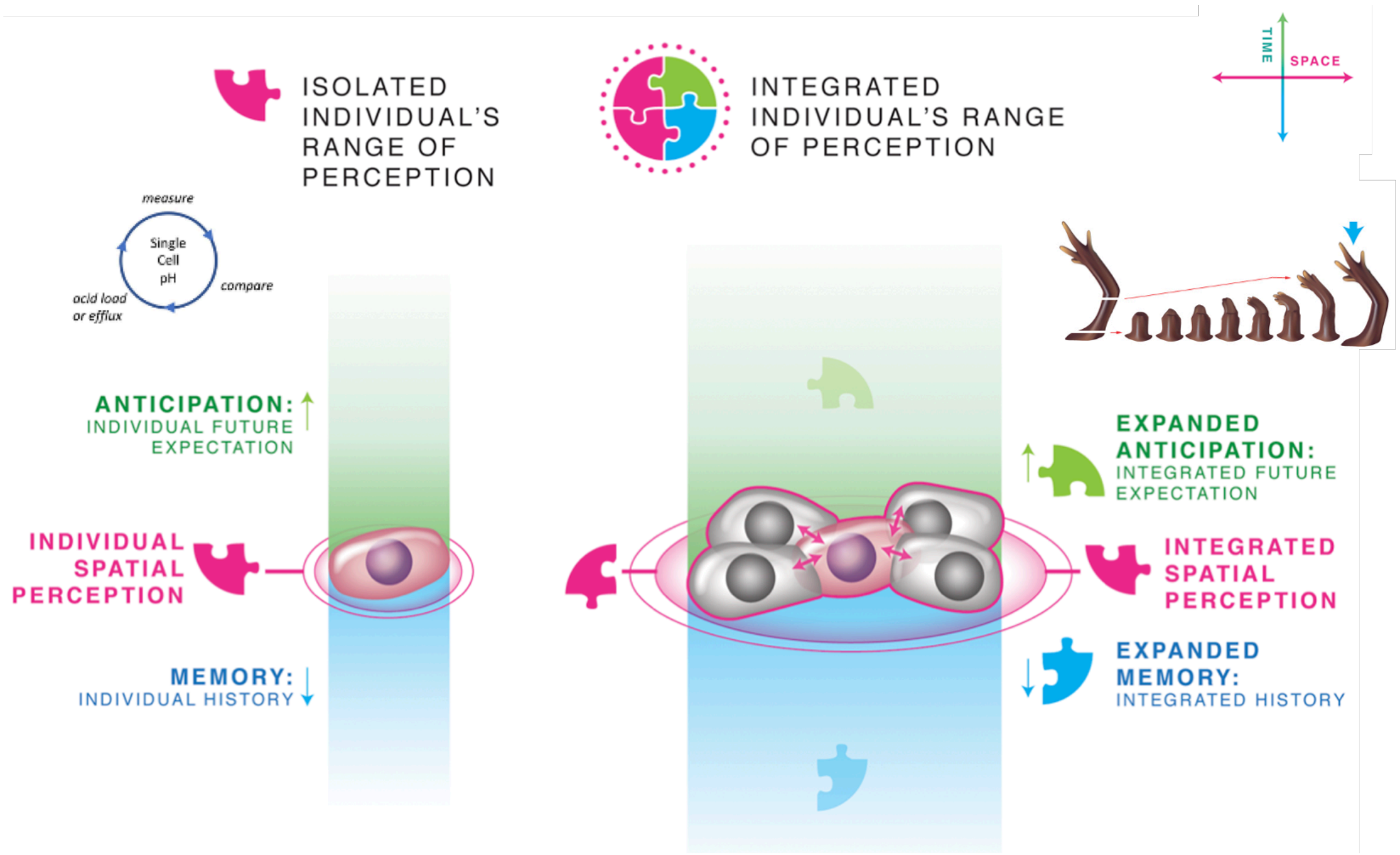

***Figure 3. A cognitive light cone model of scaling of goals.*** *The “cognitive light cone”* [19] *indicates the size, in space/time or in other spaces, of the largest goal state a system can pursue. Individual cells have tiny goals, comprising metabolic or physiological states in a very small region of space-time (microns of cell size and minutes of memory and anticipation), left panel. However, groups of cells merged together via electrophysiological sharing of information form a network that can store much larger setpoints, for example specific patterns of whole organs (a goal in anatomical*

*morphospace) and reliably work towards restoring them if deviated by injury, such as occurs in axolotls (right panel). Images courtesy of Jeremy Guay of Peregrine Creative.*

Another way to say this is that progress in living systems can be understood as the expansion of the scope and timescale over which they *predict* and *control* the environment—from the prediction and control of immediate, local effects to events that are distal in space and time, as highlighted by neuroscience theories like predictive coding, active inference, and related ones that emphasize generative modeling in the brain [21–27]. Simple living organisms, like cells, can by moving around and emitting chemicals influence the environment at the scale of a few millimeters and milliseconds (although these short-range interactions can compound to generate much wider effects). Complex creatures like us—especially when we are organized in societies—can exert control over dynamics that occur much farther away in space, including other people, and can consider events that are distal in time (e.g., recalling the past and imagining the future). One way to think about it is as the feedback between intelligence and the *radius of care* – the cognitive light cone captures the size of things the agent functionally cares about, which is an entry point into implementing valence and motivation at higher levels [28].

This expansion of capabilities for prediction and control determines a much larger space of problems that can be addressed and solved. But there is a crucial twist: the larger the space of problems, even larger the problem of *future* problems that can be addressed. Innovation creates the conditions for even more innovations, expands the space of adjacent possible [29]. Take food innovation as an example. In food innovation, a combinatorial explosion happens when the outcomes of early recipes are reused as ingredients for later ones. For example, start with just milk, flour, and heat. At first, these produce only a few simple outcomes: flat dough, heated milk, and a basic milk-flour paste. On their own, these seem like end points. But once flat dough is treated as an ingredient, it can be baked into bread, fried into flatbreads, or cut into noodles; heated milk can be cultured into yogurt, reduced into condensed milk, or combined with flour paste to form a sauce; and the milk-flour paste itself can be baked into a custard or dried and sliced for later use. Each of these new dishes can then re-enter the system as ingredients—bread becomes crumbs or sandwich bases, yogurt becomes a marinade, sauce becomes a filling—so every new outcome multiplies the number of possible future dishes. The result is not linear growth but a combinatorial explosion, where a small set of starting ingredients generates an ever-expanding space of foods because recipes recursively become ingredients for new recipes.

Yet the food innovation example remains limited in that it focuses only on new goals constructed through creative recombination of existing ones, within a fixed and predefined problem space. Biology allows for other forms of expansion of the space of possibilities, as exemplified by the examples in tadpoles, planaria, and kidney tubules mentioned in Figure 2. Additional examples include mutations that drive evolutionary dynamics, as well as genuinely novel opportunities and problems that arise from environmental change (e.g., the transition from marine to terrestrial life [30]) or encounters with novel species and life forms. All these examples highlight not just adaptation or recombination within an established problem space, but also the redefinition of the space itself, entailing new competences and boundaries, the repurposing of components for novel goals, and the reinterpretation of inherited structures in changed contexts.

The expanding landscape of problems invites organisms to continually expand their cognitive light cones and capabilities, particularly in the context of the evolutionary tug-of-war that unfolds across individuals and species. For example, if a neighboring organism becomes stronger, faster, or more efficient, others must in turn adapt and improve to remain competitive. Similarly, as an organism's actions begin to affect—or be affected by—agents operating at increasingly larger spatial or social scales, it must develop novel skills to predict and respond to these broader interactions. But it's not only the uncertainty of the external environment that drives intelligence upward: it's also the internal components, which are mutated and change on many temporal scales, requiring constant vigilance, effort, and plasticity in order to continuously align active components toward large-scale goals [31].

This dynamic implies that the recipe for intelligence cannot be static or fixed. Some species may persist for long periods by relying on established strategies, or by acquiring incremental improvements that address familiar problem spaces. However, when the structure of the problem space itself changes substantially, such adaptations are no longer sufficient. Because the landscape of problems continually expands, intelligence must support not only *flexibility*—the ability to devise new solutions within familiar contexts—but also *scalability*: the capacity to confront entirely novel problems and pursue entirely new goals in ever-expanding problem spaces.

***Scalable problem-solving in biology***

It is useful to slightly expand William James's definition, by linking intelligence not only to *flexibility* in problem-solving, but also to *scalability* and *progress* within the space of problems that can be addressed. From this perspective, living organisms are not merely capable of solving existing problems; they can also explore novel possibilities, set new goals, and invent and solve increasingly complex problems. In other words, organisms can *expand* their cognitive light cones [28], where this expansion corresponds to an extended capacity to predict and exert control over their environment (including other organisms). Primitive systems' basic drive for information (infotaxis) leads, when scaled up, to the ability to find new problems to solve. In this way, intelligence is cumulative, allowing life to continually scale its cognitive capabilities and open new regions of problem space. The distinction between a system capable of flexibly solving fixed tasks and one that actively scales its capabilities—actively seeking and devising novel problems and goals, as well as novel and creative ways to solve them—is also central to definitions of creativity, fluid and open-ended intelligence, self-improvement, meta-learning, and meta-optimization in both cognitive science and machine learning [32–35]. This extension of William James's definition is therefore not a mere semantic refinement, but defines an entirely new range of capabilities, apparent in the biological realm and desirable in machine intelligence.

For most species, problem-solving expansion occurs primarily across evolutionary timescales, while for some species it also unfolds over the course of an individual lifetime. It also enables individuals to create—and then project themselves into—entirely novel problem spaces. Whereas single cells primarily navigated physiological and metabolic spaces, their organization into large-scale collectives allowed exploration of anatomical morphospace. Living systems do not merely solve problems in order to achieve predefined goals; they can also generate new goals when existing ones become unreachable or irrelevant. This can sometimes manifest as an intrinsic drive to play and engage in curious and open-ended exploration [36,37]. As noted above, problem-solving expansion commonly emerges as a result of evolutionary arms races, but can also be facilitated by inherent properties of mathematical, not just physical or biological, dynamics [38]. In this way, successive expansions of cognitive light cones are retained and compounded over time, giving rise to an "intelligence ratchet" that progressively opens new regions of problem space [39].

On a more negative note, this expansion of cognitive scope also entails an expansion of potential *failure modes*. As biological intelligence becomes richer, it also becomes potentially more vulnerable to breakdowns at multiple levels; cognitive systems are susceptible to high-level disruptions, such as "loss of motivation" and "thoughts that break the thinker", to which simpler mechanisms are immune. In morphogenetic control, this can result in dissociative conditions such as cancer [40] and aging [41]. In more complex forms of life and cognition, these failures can manifest as maladaptive dimorphisms or as psychopathological conditions, where mechanisms that normally support inference, self-modeling, and goal-directed behavior become dysregulated [42–44]. Thus, the same architectural features that enable rich, scalable intelligence also open up new spaces for dysfunction.

Finally, while we have emphasized expansion, biology does not invariably move toward larger-scale integration, wider control, or greater complexity, and even when it does, progress can be highly nonlinear [45]. For example, transitions between unicellular and multicellular organization can occur in multiple, fragmented, and nonlinear ways. Likewise, contraction of the relevant spatial or temporal scope can sometimes be an intelligent response to changing circumstances. Coordination among collectives can generate collective competence, but it can also produce conflict, rigidity, defection, or collective failure, leading to instability and dissolution of the collective under such pressures. The above discussion of cancer illustrates this point. Cancer can be interpreted as a breakdown of organism-level coordination, but the affected cells do not thereby lose their local goal-directed capacities; rather, the scale at which goals are coordinated changes. Thus, while evolution provides numerous examples of expanding cognitive light cones, as illustrated by many transitions from simpler to more complex forms of life and organization, cognitive light cones can also contract, fragment, overlap, or become mutually incompatible. Together, these dynamics make intelligence inherently fluid, dynamic, and continually reconfigurable.

***Diverse intelligence includes machine intelligence***

The notion of diverse intelligence suggests that intelligence can be realized in many unconventional substrates, including some that may not yet have been discovered [14]. A wide range of entities can, in principle, manifest intelligence as flexible problem-solving: not only individual agents but also collectives of individuals, as well as hybrids of humans and machines; not only living organisms but also algorithms; not only brains, but potentially memories or even thoughts themselves as “thinkers,” implying that data structures can be active rather than merely passive [46–48].

Some AI systems and robots can satisfy the criteria of flexible problem-solving, despite their very different substrates and evolutionary histories compared to living organisms. Biological systems are not designed in the traditional engineering sense; instead, they evolve under stringent pressures, such as the need to sustain themselves using limited resources, to continuously interact with a dynamic environment, and to cooperate and compete with other organisms. AI systems, by contrast, are largely designed by humans—we specify their architectures and curate their training data—and are subject to different “pressures” than those shaping biological evolution. Nevertheless, defining intelligence as flexible problem-solving does not exclude a priori artificial systems.

When discussing modern artificial or machine “intelligence”, large scale systems – such as large language or large multimodal systems, or systems capable to address challenging games – immediately come to mind. However, even much humbler systems, such as sorting algorithms implemented in multi-agent settings, might comply with William James’ definition by displaying some forms of flexibility when solving their tasks, for instance by navigating around defects or unexpected constraints, or exhibiting other competencies recognizable to behavioral scientists toward goals not explicitly specified in the algorithm [49]. Yet this does not mean that we should automatically label as intelligent any form of computation. A simple piece of code—for example, a for loop that iterates over a vector of numbers—may not exhibit the flexibility implied by William James’s definition, but an empirical approach is necessary because such properties can occur in systems whose algorithm (or, in the case of biological organisms, whose mechanisms) does not explicitly encode it. Tools must be developed to test a wide range of systems for unexpected flexibility and novel goals in unconventional spaces, to establish rigorous results for adjusting our insufficient intuitions.

We can also move beyond the observation that diverse forms of intelligence—including machines—navigate spaces of problems to be solved. More speculatively, it is possible that these systems occupy a *shared* problem space with a relatively narrow landscape of viable solutions, such that agents starting from very different substrates, evolutionary histories, or learning algorithms may nevertheless converge on sufficiently similar intelligent behaviors, representations, and functional manifolds. In nature, this is reflected in convergent evolution: similar solutions, such as the repeated emergence of eyes, have arisen independently across lineages. Similarly, in artificial systems, different architectures, training regimes, and modalities—ranging from text-based language models to multimodal models incorporating video—appear to converge on comparable internal geometries of representations [50]. These convergences might occur for relatively trivial reasons (e.g., if multiple systems share training data or architectural constraints), but they might also arise, at least in part, from deeper principles: systems may be constrained or enabled by the same statistical structures and problem spaces, even when their histories, materials, and algorithms differ. In other words, one might speculatively argue that multiple systems facing the same natural statistics—for example, the distribution of natural scenes encountered by systems developing visual capabilities—and similar problems and constraints would be implicitly constrained in the range of solutions and adaptations they could develop to succeed. If living organisms and artificial agents share – at least in part – overlapping spaces of problems and solutions, it becomes particularly valuable to study the solutions discovered by biology: how they were discovered, organized, and realized over evolutionary and developmental time. This is not necessarily to copy nature's solutions, but to understand the underlying principles that make flexible and scalable intelligence possible.

In summary, we have introduced a theoretical perspective—rooted in the notion of diverse intelligence and William James's definition of intelligence as flexible problem-solving—that highlights the remarkable diversity of forms and manifestations of intelligence found in biology. Biological evolution has discovered a recipe for *flexible* and *scalable* intelligence: the capacity to pursue goals in many conditions and the continuous expansion of organisms' capabilities, or *cognitive light cones*. One might even argue that the evolution of life itself represents a remarkable instance of problem-solving: biology has discovered effective ways to generate diverse and increasingly sophisticated forms of intelligence, all while operating under fundamental constraints (e.g., the laws of physics) and evolutionary pressures (e.g., survival, efficient energy use). Much remains to be investigated about the level, along a cybernetic hierarchy [13], of evolutionary search algorithms carried out over a multiscale agential substrates [39]. We have also argued that the notion of diverse intelligence encompasses not only biological intelligence but also machine intelligence, providing a common framework for understanding the capabilities of both—in terms of the problems they can solve flexibly—and the (shared) space of possible solutions to those problems.

Crucially, many of the feats routinely observed in biological systems—such as flexibility (the capability to pursue autonomously-set goals in many conditions and to generalize "out of distribution" to implement adaptive outcomes when internal parts or environmental changes make it impossible to reach species-specific solutions as their ancestors did) and scalability (continual learning and self-improvement, autonomous exploration of novel possibilities and goals)—remain largely unparalleled in current AI systems. How does biology achieve these impressive capabilities? And what lessons might they offer for the development of more capable machine intelligence?

**How does biology support flexible and scalable intelligence? Five design principles**

In addressing "the problem of life," biological systems leveraged five fundamental design principles – autonomy, growth through multiscale self-assemblage of active components, continuous reconstruction of capabilities, exploitation of physical and embodied constraints, and pervasive signaling enabling self-organization and top-down control from goals – which we illustrate here and from which we extract potential lessons for the development of machine intelligence.

***Living organisms are "natural born autonomous"***

Living organisms, from simple cells to complex plants and animals, are inherently autonomous. Unlike artificial systems, they are not programmed or supplied with external data and goals. Instead, from the outset, they must continuously act (e.g., move around, scan the environment, forage for food), make autonomous decisions to fulfill goals of varying complexity, and learn from interactions, all while maintaining functionality over extended periods. They must decide where to look, set and pursue their own goals, autonomously determining their own agenda to manage internal states and environmental cues.

Furthermore, living organisms are *autopoietic*: they must generate and maintain the conditions necessary for their own existence and integrity [51]. This implies that the organism is both self-constructing and self-directing, continuously orchestrating its metabolic, behavioral, and cognitive processes to sustain itself while producing and maintaining the very components that realize it.

In this sense, autonomy, goal-directedness and autopoiesis are intrinsic properties of life, emerging naturally from the interplay between the organism's internal dynamics and its interactions with the environment. Autonomy manifests itself in any domain: *active* sensing beyond passive sensory processing, *active* (and purposive) control of behavior beyond fixed stimulus-response, *active* learning beyond the passive ingestion of training data (as action solicits useful targets for learning), *active* control of embodied and physiological processes, such as energy management, metabolism, shape, and growth, *active* goal setting, implying autonomous and creative generation of new goals and ways to solve them, etc.

Living beings exist in a time- and energy-constrained environment, so they are forced to learn to coarse-grain signals (they do not have the luxury of being micro-reductionists or Laplacean Daemons). This in turn drives generalization and eventually agentic modeling (which, in advanced cases, can be turned inwards to generate a sense of Self). Not only must organisms independently make sense of (interpret) the outside world and their own internal information media, but so do all of their parts. Living materials have agendas and autonomy at every scale, all of which are trying to communicate with, hack, and behavior-shape each other.

For many advanced organisms like us, the first act of autonomy—or perhaps proto "free will"—occurs during embryogenesis, when the developing organism begins to act as a coherent whole. In this process, "being (or becoming) you" emerges as a causal entity, liberated from the tyranny of molecular-level determinants. The organism becomes guided by a larger, self-generated story: a model of what it could become in the future, where it can act within anatomical space, and what possibilities lie ahead. "An embryo" is a single thing in the same sense as a brain is a single thing, despite both of them being made of a huge number of cells; what is being counted there is alignment: the ability of their parts to merge into a collective that has memories and preferences that none of the parts have and that will pursue novel goals as a unified whole that none of the parts can achieve individually. In Aristotelian terms, this represents a distinction between material cause (the molecular constituents) and final cause (the self-generated purpose or goal) [52–57]. Crucially, the final cause (the goal) is internally generated and maintained by the organism itself, providing a notion of teleology that is not mysterious but firmly grounded in the organism's own biological processes and now characterized biophysical mechanisms that implement goal-directed activity [58,59].

A crucial requirement for this and other acts of autonomy is that an organism maintains its identity and integrity—distinguishing itself from the environment, which is subject to thermodynamic forces and dissipation—while remaining sensitive to it through sensors and capable of acting upon it via effectors. One way to formalize this distinction is through the concept of a Markov blanket [23,60]: a (statistical) boundary that mediates interactions between an organism's internal states and the external environment, with the interface realized through the perception–action cycle (Figure 4).

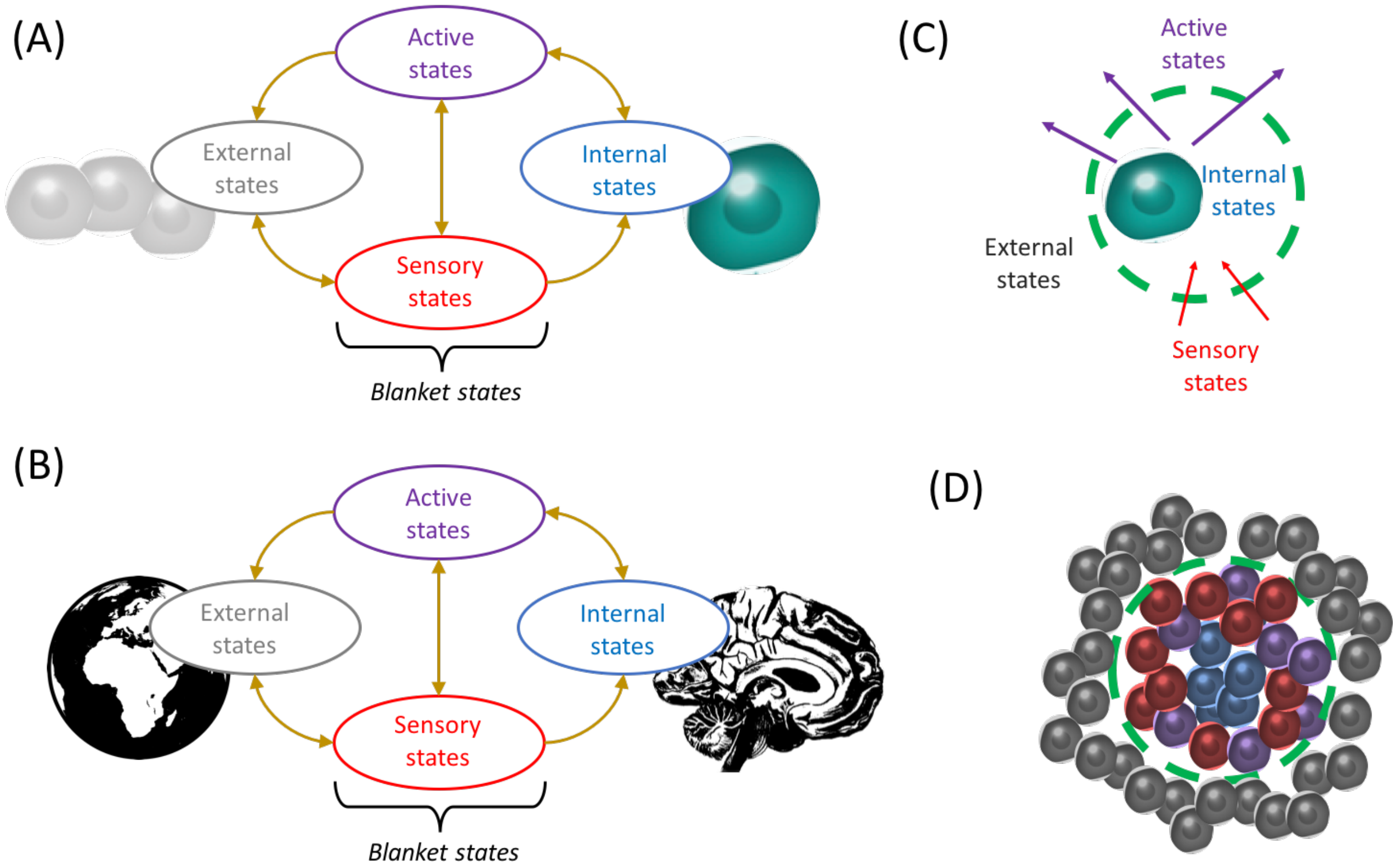

***Figure 4. Autonomy and Markov Blankets.*** *(A-B) Single-level autonomy. Autonomous exchanges between an organism and its external environment, which may include other organisms. Examples span multiple levels of complexity: a single cell actively regulates its internal states relative to its immediate environment, while advanced organisms, such as mammals, maintain a brain that integrates sensory inputs and guides actions to achieve goals over longer timescales and larger spatial scales. These interactions illustrate how Markov blankets define a statistical boundary between internal (blue) and external (grey) states, mediated by sensory (red) and active (violet) states, effectively mediating action-perception loops. (C-D) Nested Markov blankets. Examples of hierarchical or nested Markov blankets, illustrating the scaling of agency and control. Individual cells maintain their own blankets (dotted green circle around a single cell) to regulate internal states, yet collectively form the tissues and organs of multicellular organisms, which are themselves encompassed by a larger organism-level Markov blanket (dotted green circle around multiple cells, playing the roles of internal, sensory and active states, shown in blue, red and violet, respectively), mediating interactions with other cells playing the role of external states (grey cells). At higher levels, social groups or colonies could similarly be conceptualized as having emergent blankets that regulate collective behavior. Nested blankets allow each level to retain a degree of autonomy while being constrained and coordinated by larger-scale dynamics, supporting robustness, flexible problem-solving, and the expansion of cognitive and behavioral capabilities.*

The Markov Blanket provides a statistical description of the organism's ability to *model* and exert *control* over the environment, therefore determining the span of their "cognitive light cone". While processes such as autopoiesis, self-organization, and the active maintenance of internal viability states generate boundaries between the organism and its environment, the Markov Blanket formally describes these boundaries as a statistical partition between internal, sensory, active, and external variables (see also [61,62] for a discussion of how probabilistic dependencies and possible Markov blankets could be derived from principles of self-organization and organizational closure). By identifying a boundary on the region of a problem space which an agent tries to manage (the size of goals it actively cares about), it in effect provides a statistical measure of the scale of its Self – the boundary between itself and the Outside World [19]. In the Markov Blanket account, internal states act as a (generative) model of external dynamics, informed by sensory observations, which are in turn autonomously and purposefully solicited through action. Actions are driven (implicitly or explicitly) by the goal of maintaining certain internal states, which play the role of goal states, within viable ranges (e.g., for simple organisms, homeostatic goals related to viable ranges of temperature or chemical concentrations). Reaching these goals is a primary objective, and it is achieved using cybernetic-like mechanisms, consisting of checking any discrepancy (or prediction error) between goal states and sensations and correcting it by acting if necessary. The capability to organize behavior around goals (or "set points" in cybernetics, or "higher-level priors" in predictive processing), rather than fixed stimulus–response rules, ensures flexibility, since the same goal can be pursued in variable ways [63–65]. While this notion immediately suggests homeostatic mechanisms, goal achievement can also be (and in living organisms, often is) achieved through *allostatic* mechanisms—which anticipate discrepancies from goals and compensate for them before they occur [66,67].

Crucially, the goals that are fundamental for survival are central in brain architectures [68–70] and need to be protected from external perturbations (i.e., to be relatively resistant to updating in light of sensory evidence, unlike the typical beliefs of organisms) [23]. The Markov blanket – and the continuous activity required to preserve these goals – endow the organism with a rudimentary form of "self" that is stubbornly maintained and persists over time, so that the organism remains distinct from, and not entirely governed by, environmental fluctuations. In advanced animals like humans, goals are organized hierarchically—from low-level drives related to homeostasis and self-preservation to increasingly distal, abstract, and cognitive goals. These hierarchies are supported by sophisticated mechanisms of cognitive control and self-regulation that allow organisms to prioritize among competing goals [71,72]. Remarkably, this can include the capacity to suppress or even sacrifice immediate self-preservation in favor of higher-order aims, such as long-term commitments, social values, or the maintenance of a coherent, meaningful narrative about the self over time.

Goal-directed behavior and active sensing require a *model* of action consequences and the environment—which in machine learning is referred to in various ways (with distinctions that are not crucial here), e.g., a generative model, predictive model, or world model. The model is not a copy of external reality. Metabolic and temporal constraints necessitate coarse-graining, forcing living systems to build simplified, predictive models of the world [73]. But it is not just a matter of limited resources: what matters most is that the model is good enough to steer adaptive action (which is primary), rather than to replicate external dynamics in the brain, let alone to provide a "rich" or fully "veridical" representation of them (which is ancillary) [74].

Another way to illustrate the intimate link between internal models and control is through the "good regulator" theorem, which, in simple terms, states that any system or living organism capable of adaptive control over its environment must have, or embody, a model of that environment [75]. Applied to biological organisms, this notion can be interpreted either instrumentally—as a way of describing the organism's behavior through the lens of a predictive model—or, alternatively, as a claim about the organism's internal organization and functioning (necessarily subject to empirical scrutiny). Even in the latter case, however, the notion of a model should be understood as spanning a continuum. At one end, it may refer to a full-fledged internal model, in which the organism's organization and dynamics map to variables required for predictive processing and control, akin to classical notions of small-scale models [76] or modern AI notions of world models encoding states and their transitions conditioned on an agent's actions [74,77,78]. At the other, it may refer to more minimal models whose mapping with such variables is more implicit, either because they are in part absorbed by the system's embodiment (a notion of "being" or "embodying" rather than "having" a model) or because they are distributed across the system's dynamics and behavioral policy; indeed, aspects of a model—specifically, the environment's transition dynamics—can be present "in disguise" even in systems that might otherwise be considered model-free (essentially stimulus–response), as they can be implicitly encoded in and extracted from the agent's behavioral policies [79].

In simple living organisms, the internal model – being it full-fledged, more implicit or "embodied" – needs only to support simple strategies. In bacteria, the internal model supports decisions between swimming in a straight line ("runs") and selecting a new random direction ("tumbles"), and it is carefully designed to guide the organism toward areas richer in nutrients by increasing the probability of tumbling when conditions worsen. In C. elegans, a dynamical internal model describes—or more precisely, prescribes—largely preconfigured transition dynamics between locomotion behaviors, with a few bifurcation points that embody simple decisions [80]. These relatively simple internal models, which focus far more on actions and action sequences than on creating detailed representations of the external world, are sufficient for organisms to be highly successful from an evolutionary viewpoint.

In more complex organisms, internal models can be far more sophisticated, incorporating detailed sequential and predictive dynamics. Nevertheless, one can argue that the most important predictions—whether explicit or implicit in the model dynamics—concern the consequences of action, as these are what enable living organisms to achieve their goals, above and beyond merely understanding the environment. For example, whisker movements in rodents can be interpreted in terms of generative models that capture the sensory (primarily tactile) consequences of whisker motion. Animals can then leverage these models for active sensing, recognizing objects based on the distinct sensorimotor contingencies elicited when objects are "probed" through whisking [81]. But importantly, these kinds of processes are already happening in single cells and cellular collectives [82,83].

Along this continuum, successive evolutionary transitions can be understood as an expansion in the structure and depth of organisms' generative models [84]. From single cells to multicellular organisms, and from animals with relatively simple nervous systems, such as C. elegans, to species with increasingly complex brains, organisms have progressively developed more expressive models of the world – from relatively flat, reactive architectures to hierarchical and temporally deep models encoding regularities across the past, present, and future – and, in parallel, richer models of the self, spanning multiple levels, from representations of bodily structure and internal dynamics to abstract, narrative models that support a continuous sense of identity across time [85,86] as well as models for cooperative social interaction [87–89]. Progress in the "cognitive light cone" of living systems can therefore be understood as the expansion of the scope and timescale over which organisms model and control their environment—from immediate, local effects to events that are distal in space and time. This process is incremental: existing predictive and regulatory capabilities scaffold the emergence of more sophisticated ones. In this view, advanced cognition arises as a phylogenetic extension of predictive control and allostatic regulation mechanisms, which themselves build on much older capabilities already present in simple life forms, i.e., a phylogenetic refinement of capabilities [90,91]. Work is on-going to understand whether these dynamics already exist in chemical networks below the cell level, which exhibit simple forms of learning [92–95] and causal emergence (higher levels of control) linked to their learning history [96].

Within the generative models that living organisms use, action-based predictions can operate over short or long timescales and take different forms, all of which are important. Proprioceptive predictions (e.g., anticipating that one's finger will be raised) guide adaptive motor control and active sensing in active inference: they generate prediction errors when the expected state is not yet realized, which in turn drive actions to minimize these errors (e.g., raising the finger). Sensory predictions, often implemented via corollary discharges, allow organisms to distinguish self-generated stimuli from external events, supporting accurate perception and interaction with the environment. Interoceptive predictions evaluate whether a course of action is beneficial or harmful to the organism (e.g., will it increase the chance of obtaining food?).

In more advanced animals, generative models become richer and more elaborated. Humans, for example, possess factorized generative models for "what" and "where" visual streams, as well as rich cognitive maps supporting navigation in both physical and abstract conceptual spaces. These models can also "detach" from the immediate here and now, enabling planning, imagination, and the anticipation of distal outcomes. This elaboration represents a natural continuation of the incremental expansion described above, where hierarchical and temporally deep models build upon simpler, evolutionarily older capabilities to support increasingly flexible and adaptive goal-directed behavior.

Importantly, autonomy extends beyond perception and action to encompass learning. Organisms learn autonomously, continuously, and interactively from experience, using their actions to satisfy existing goals and to generate novel ones in ever-changing environments. Crucially, this learning is active: life forms do not passively observe the environment but engage in exploratory behaviors that elicit informative feedback, allowing them to refine their models, identify the strategies that reliably work and generate entirely new solutions. Through these interactions, organisms acquire generative models of the physical world and develop competencies that support goal-directed activity, forming the foundation for the flexible and scalable intelligence observed across biological systems. Learning is largely self-supervised, guided by internal drives and cognitive goals, although social species may also benefit from cultural and social instruction.

These internal supervisory mechanisms are typically generic, designed to optimize for the future: evolution cannot predict all novel situations—such as the dangers of driving too fast—but it can produce general learning algorithms that adaptively create and refine goals to navigate unforeseen circumstances. This form of autonomous learning requires balancing pragmatic goals—those that maintain homeostasis and meet basic survival needs—with epistemic goals—those that reduce uncertainty, promote curiosity, and foster creative exploration. The continuous drive to expand capabilities, rather than simply reuse existing solutions, lies at the core of behavioral flexibility and adaptive, scalable intelligence [23].

Summing up, living beings exemplify intelligence as inherently autonomous, goal-directed, and actively learning. From single cells to complex animals, life continuously models, predicts, and interacts with the environment, extending its agency across space, time, and hierarchical scales. Autonomy, generative modeling, and the protection of core goals ensure both persistence of self and the flexibility to adapt to novel challenges. These processes highlight that scalable intelligence emerges not from pre-specified solutions, but from self-directed, embodied, and actively constructed interactions with the world.

***Organisms' bodies and cognitive abilities grow through the multiscale self-assemblage of active components***

A key cornerstone of life is the scaling of problem-solving capabilities—traversing increasingly complex problem spaces. This expansion supports an increase in both cognitive capabilities and agency, allowing organisms to coordinate behaviors and predictions across multiple levels of organization while retaining local autonomy. How is this possible?

Crucially, bodies and cognition emerge through continuous processes of growth, self-organization, and self-assembly, scaffolded by prior iterations. Complex living organisms are composed of active components that are themselves agentic—simpler organisms acting collectively, rather than inert parts. Multicellular organisms, for instance, are made of cells that are autonomous agents in their own right. At an even smaller scale, molecules implement a sensing-decision-actuation loop across environmental contexts, displaying a primitive form of adaptive competence [97,98], providing a rich substrate for intracellular activities involving proteins, DNA, and RNA [99,100]. This occurs not only at the level of individual molecules, but also in molecular networks, such as in cases of probabilistic inference in simple biochemical circuits [101,102]. This principle extends across developmental and evolutionary scales: intelligence is fundamentally multiscale and collective, with each level of organization contributing to problem-solving capacities and defining the solution manifold available to the organism. Living beings at each level continuously exert effort over their active parts, seeking to control their behavior lest they veer off to focus on their own local goal states [103]. The constant, inescapable need to align and behavior-shape one's internal components is shaped by efficiency pressures to take advantage of proto-cognitive capacities of those parts.

At the cognitive level, living systems build robust layers of competence. Novel, more sophisticated solutions are added atop existing strategies, coordinating with rather than replacing them. Consider the defensive brain circuits of organisms [104]: they resemble layers of fortifications, analogous to the defensive walls of a medieval castle, protecting against threats that vary in distance and immediacy. Each layer maintains some autonomy, yet all layers coordinate to ensure robustness. This principle extends beyond brain systems; animal bodies coordinate multiple internal systems—circulatory, excretory, microbiota regulation, adaptive immunity, and sensory-motor systems—without centralized control. Such distributed, modular architectures ensure that failures or perturbations at one level do not compromise the organism as a whole (Figure 5).

At the neural level, early brain designs present in our evolutionary ancestors provide a scaffold for later evolutionary developments. The sophisticated predictive and control abilities of evolutionary recent organisms like us reuse and gradually elaborate simpler predictive and control abilities that were already present in our early evolutionary ancestors; for example, by expanding simple generative models into hierarchical and temporally deep generative models (that afford hierarchical inference, long-term prediction and imagination), while typically preserving rather than completely replacing previous functionalities [84]. Even those that we consider our more complex cognitive skills—such as planning, mindreading, and language—typically reuse and develop on top of simpler sensorimotor and predictive mechanisms. This reflects exaptation: higher-order cognitive abilities often repurpose neural mechanisms originally selected for other functions.

Importantly, the principle of self-assemblage extends to learning. Living material is trainable from the very bottom: even molecular networks exhibit forms of learning. Organisms learn across scales, from cellular signaling networks to whole-brain architectures, acquiring competencies that support adaptive, goal-directed behavior. Self-assembly is autonomous and spontaneous, echoing Kauffman's insight that living systems inherently self-organize and can spontaneously increase order, providing a foundation for progressively more complex and capable organisms [108].

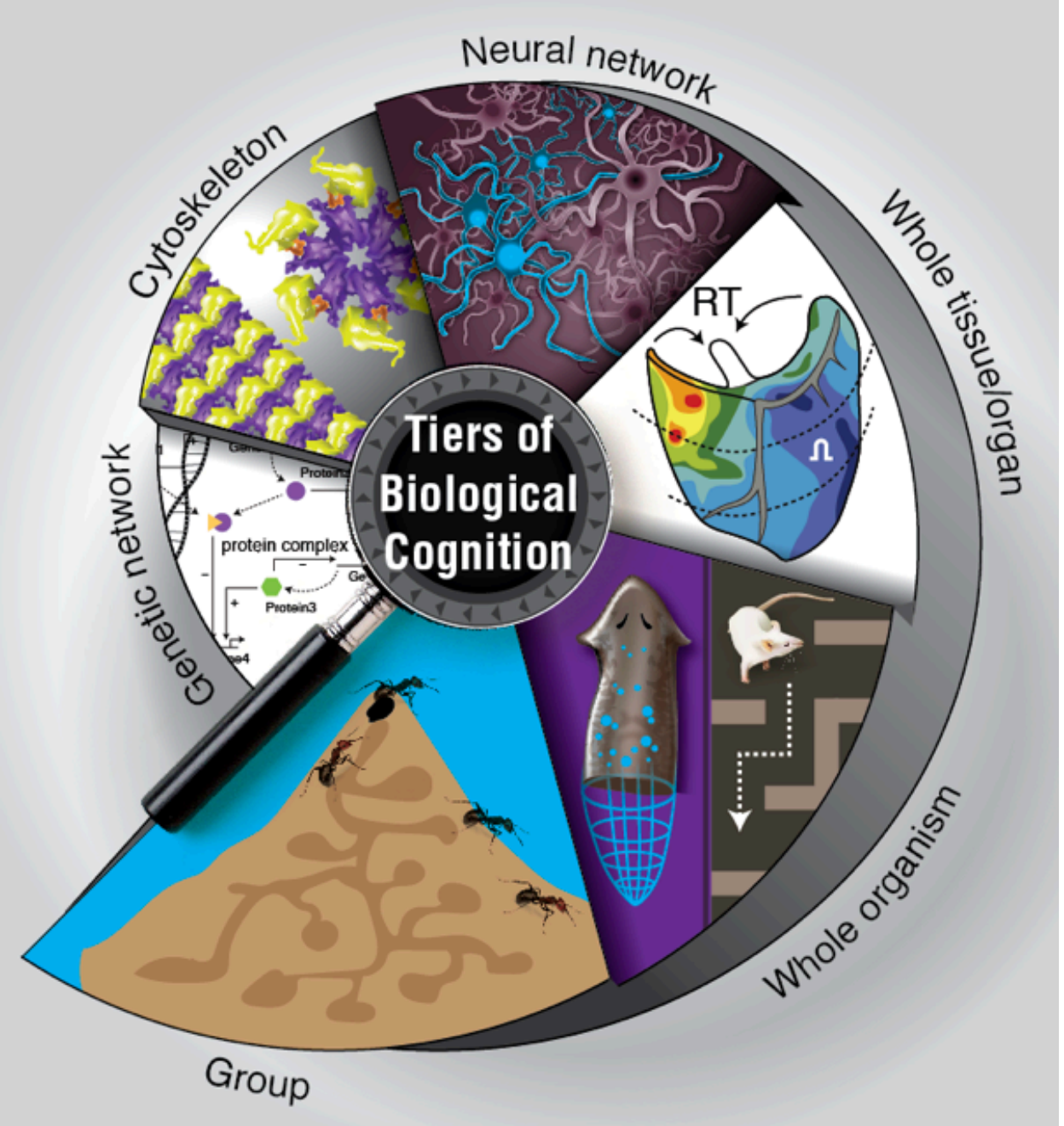

***Figure 5. Multiscale competency architecture of biology****. Biological systems exhibit active problem-solving at all scales – from memory in molecular networks* [92,93] *to physiological and anatomical plasticity of cellular and tissue-level scales, in addition to conventional organism-level behavior and increasingly-recognized collective intelligence of groups and swarms* [105–107].

The expansion of capabilities can be formalized in terms of *nested* Markov blankets (Figures 4-5). In this perspective, pursuing more sophisticated goals and plans corresponds to an expansion and nesting of Markov blankets, from boundaries around single cells to larger blankets that afford acting "at a distance" across tissues, organ systems, and whole organisms. Individual cells maintain blankets that regulate internal states relative to local environmental conditions, allowing autonomous behavior and homeostasis. These cellular blankets are nested within organism-level blankets, which coordinate prediction and control across tissues, organ systems, and ecological interactions. Importantly, the agency of the lower-level blankets is not eliminated: cells continue to act according to their own goals, but these actions are constrained and integrated within higher-level dynamics. The highest-level goals propagate downward, guiding interactions among parts while preserving local autonomy.

This hierarchical nesting illustrates a core principle of life: scaling of cognitive capabilities and agency. It allows simpler, local adaptive processes to be coordinated into higher-order behaviors, achieving both robustness and flexibility. The organism's overall cognitive light cone expands across levels, while subcomponents retain essential problem-solving capacities. This multiscale, self-assembling architecture offers a biologically inspired blueprint for artificial intelligence, suggesting that future systems could integrate multiple layers of self-modeling, prediction, and control, preserving local autonomy while enabling coordinated global behavior.

Summing up, living organisms achieve scalable intelligence through the continuous self-assembly of active, agentic components across multiple levels of organization. From molecular processes to cells to tissues to whole organisms, nested hierarchies of control and prediction allow local autonomy while coordinating higher-order goals. By building new capabilities atop existing ones and repurposing previous mechanisms, life achieves both robustness and flexibility, enabling organisms to navigate ever more complex problem spaces and expand their cognitive and behavioral reach. Yet, nesting provides only one possible statistical description of multiscale dependencies, which can take diverse forms and does not necessarily imply progression toward wider agency. Each expansion also introduces potential failure modes: while lower- and higher-level regulatory structures can cooperate, they can also compete, become disconnected, or pursue incompatible goals (see below for cancer as an example of such a disconnect between levels of organization).

***Living organisms continuously rebuild themselves – and achieve adaptation through change***

> "If we want things to stay as they are, things will have to change" *Giuseppe Tomasi di Lampedusa, The Leopard.*

Another crucial mechanism underlying the scalability of biological intelligence is *continuous rebuilding*. Living organisms do not merely adjust parameters within a fixed architecture; instead, they persist by continuously reconstructing themselves, sometimes gradually and sometimes radically, while preserving or even expanding functional capabilities. Change is not an anomaly to be resisted but the default condition under which life operates. The environment in which organisms (as species and as individuals) live constantly changes, in partially unpredictable ways. But adaptation through rebuilding is not driven solely by environmental change; it is also necessitated by the intrinsic unreliability of biological components. Living systems are composed of parts that are unreliable: they age, mutate, die, and are replaced.

This rebuilding occurs at multiple timescales and organizational levels. Organisms continuously self-repair damaged tissues, remodel physiological systems, and reconfigure neural and molecular signaling circuits. In more dramatic cases, rebuilding takes the form of metamorphosis, where organisms undergo profound morphological and functional transformations while retaining, and sometimes enhancing, cognitive and behavioral capacities [109,110]. At the developmental and evolutionary timescales, changes can be even more dramatic. Importantly, change does not imply erasure: existing competencies are often conserved, scaffolded, and extended, forming a ratchet-like process in which new capabilities accumulate on top of old ones rather than replacing them. A critical feature of living organisms is their capacity to adapt to change: some are *resilient*, maintaining their capabilities in the face of perturbations, while others can be even *antifragile*, increasing their capabilities by exploiting internal and environmental variability and perturbations as opportunities for adaptation [111].

Key to this capability is the fact that the evolutionary "master program" responsible for generating organisms does not transmit fixed solutions, but rather general learning strategies and inductive biases that enable organisms to cope with novel, unforeseen problem spaces – in other words, to improvise. At the beginning of life, organisms do not inherit fully specified solutions from their ancestors but rather a compressed set of prompts—genetic instructions that must be interpreted, elaborated, and instantiated anew during development in each individual. This developmental "bowtie" architecture, in which minimal inherited information fans out into highly complex phenotypes, thereby forces organisms to become adept at flexible, creative reconstruction rather than the rigid execution of pre-specified programs (Figure 6).

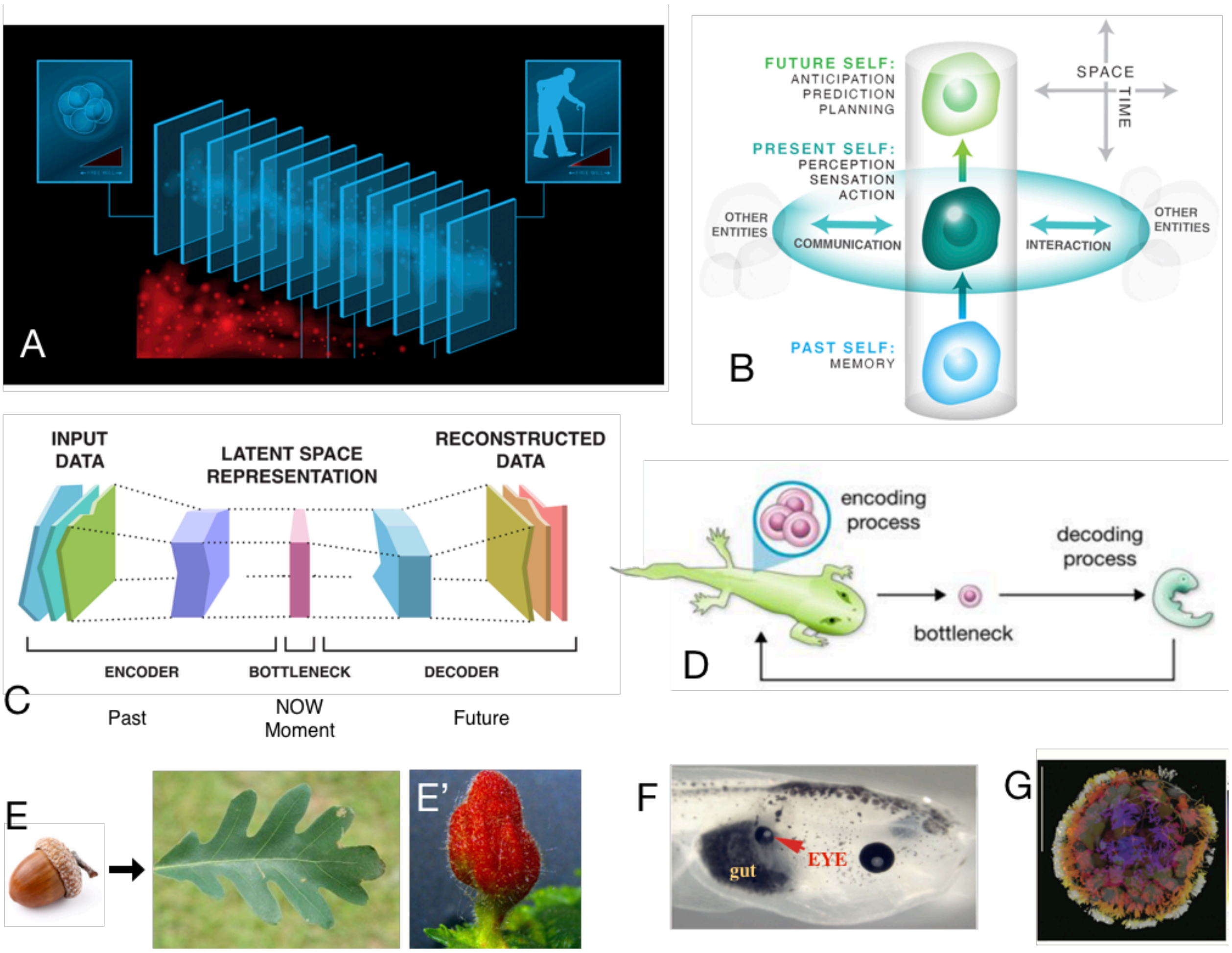

***Figure 6. Continuous rebuilding and the "bowtie" architecture**. (A) Autonomous cognitive agents, such as biologicals, do not have access to the past – they consist of a series of Selflets, each of which must interpret the biophysically-mediated memory engrams in their brain and body to continuously construct (not just restore, but actively build) a model of themselves and of the outside world. (B) In other words, memories are messages from a past Self, and like all signals, can be interpreted in whatever way is most adaptive at the time. While our current engineered systems are built on a reliable substrate and we, the users, interpret their physical states and imbue them with meaning, biological forms are responsible for knowing what their memory structures mean. (C) Importantly for the agent, residing at the ever-moving NOW moment between history and future possibilities, the process of compressing instances (experiences) into a generalized memory discards a lot of information. Memory storage and retrieval is actually a creative, improvisational process because redundancies are removed from the data (the past), and the decoding of the memory must involve bringing novel meaning to them that cannot be derived entirely from the materials. This continuous rewriting, remapping, and reinterpretation of events is a hallmark of (constructive) episodic memory [112,113] and can be metaphorically considered a form of confabulation that, contrary to its common negative meaning, is adaptive because it allows creative recombination and the generation of novel interpretations* [114]. *(D) The same bottleneck exists for morphogenesis: each organism receives not the anatomy of her ancestors, but a highly compressed parts list (the genome, encoding proteins, not organ shapes) which it is then responsible for interpreting in current circumstances to produce an adaptive outcome (which may or may not be the species-specific default). This process, of maximizing adaptive saliency, not fidelity, of past memories is not only a feature of neural systems (which have no nondestructive reads, only reconstructions of memories), but of their ancestors – development. The guaranteed lability and unreliability of the genome and other hereditary materials* [115], *requires morphogenesis to treat the genome as an input and use problem-solving processes, not*

*hardwired instruction, to turn it into a functional form* [31]. *This flexibility is why the oak genome in acorns can be prompted (by wasp parasites' signals) to build a gall instead of a flat green leaf (E, E'), why other regions of the body in tadpoles can be bioelectrically prompted to build a complete eye (F)* [116]*, and why adult human cells can form active constructs called Anthrobots despite their normal Homo sapiens genome (G)* [117]. *Panels A-D courtesy of Jeremy Guay of Peregrine Creative. Other panels taken with permission from the references indicated.*

The same principle applies to cognition. Learning in advanced animals is rarely about fixed responses: it is most often about flexible schemas to face novel situations and meta-learning strategies, i.e., "learning to learn" [118]. Memory in living systems is not a static storage device but a constructive process. Episodic memories are actively rebuilt upon recall, enabling organisms to reinterpret past experiences in light of current goals and contexts. The brain is characterized by significant neuroplasticity at the network level, and neural representations themselves are transient: for example, hippocampal place cells dynamically drift and continuously re-create spatial fields, rather than preserving immutable maps. Through such processes, organisms effectively shape their futures, using fragmentary traces and internal prompts from the past to reconstruct meaning, guide action, and adapt to novel goals. Cognition privileges solutions that work for the future—both in a future that can be anticipated and in a future that cannot and therefore requires the flexible reconstruction of strategies and the reassembly of competence.

Rebuilding is also intimately linked to mortality. Brains are not inherited ready-made; instead, each new organism begins with a powerful brain design that supports flexible learning but must be trained through interaction with the world. This is likely achieved by a flexible neural architecture capable of spontaneously generating a repertoire of internal (sequential) dynamics that are initially "meaningless," but acquire meaning when they become coupled to the environment through the animal's actions [119]. This principle of brain design exemplifies that evolution favors scalable mechanisms, improvable through lived interaction, over hard-coded solutions. Evolution cannot anticipate future contingencies—such as the cognitive and social consequences of modern technologies—but it can shape organisms that learn, adapt, and generate new goals when old ones become maladaptive. The relevant state space is itself continually shifting, requiring forms of intelligence that are plastic and (re)constructive rather than fixed.

Despite this pervasive change, living organisms maintain a persistent sense of selfhood over time. As in the Ship of Theseus, whose pieces are continuously replaced while the ship remains the same, persistence does not arise from material continuity but from functional and narrative continuity. Living organisms stabilize identity by coordinating rebuilding processes across scales. They effectively *communicate with their past selves* through memories, which act as messages guiding interpretation and action, and *with their future selves* through habits and niche construction, where current actions shape the conditions their future selves will encounter. This persistence is enabled by the maintenance of Markov blankets, which preserve a stable boundary of agency and protect core goals from being overwritten by transient sensory fluctuations. In doing so, organisms preserve agency while remaining open to transformation.

Summing up, continuous rebuilding provides a powerful route to scalable intelligence. By reconstructing bodies, models, and memories under changing constraints, living organisms transform instability into a resource. This intelligence ratchet, operating across evolution, development, and cognition, allows life to expand its cognitive light cone not despite change, but by exploiting it to generate new competencies, explore novel problem spaces, and sustain agency in the face of uncertainty.

### *Living organisms consistently exploit embodied and physical principles*

Cognition is fundamentally embodied and constrained by physical principles. The development of the body, of neural and cognitive capabilities are grounded in the physical substrate that supports them (and in addition, they are tightly coupled, co-evolving and co-developing). Crucially, constraints imposed by chemistry, physics, and energetics—such as energy maintenance, real-time processing, and limited material resources—are not merely limitations to overcome. Instead, living systems consistently *exploit* these constraints to structure behavior, shape and regularize computations, and generate adaptive solutions that would be infeasible in unconstrained, freeform systems.

At multiple scales, the physical medium itself embodies computational regularities. In morphogenesis, bioelectrical fields constrain cellular behavior, while facilitating tissue- and organ-level competencies, producing reproducible patterns and structures while effectively implementing generative rules directly in the physical substrate. In the brain, dynamical events such as oscillatory patterns, rhythms, and traveling waves encode and propagate information efficiently, enabling cognitive control and possibly forms of analog computation, where the medium itself rather than a digital system performs computation [120–122]. These constraints reduce the arbitrariness of computations: the laws of physics and chemistry create a natural manifold of plausible solutions, making adaptive behavior and development more reliable and energetically efficient.

Other physical constraints are similarly exploited in behavior. For example, at the molecular level, protein networks, cytoskeletal elements, and reaction-diffusion systems exploit chemical and mechanical constraints to perform "computation" that is robust, parallel, and embedded in the material medium. Animals must manage energy budgets and biomechanics when moving or foraging; neural and muscular systems co-develop with the body to ensure energetically efficient solutions, rather than relying on trial-and-error computation in isolation.

Living organisms also consistently *offload* and *externalize* parts of their cognitive abilities, effectively extending their problem-solving capacity beyond the boundaries of their own bodies. Even in simple life forms, externalization takes many forms: cells and bacteria, for instance, create chemical gradients or modify their microenvironments, leaving traces that can be used by themselves or others to guide behavior [123]. At the level of biomechanics, passive dynamics provide striking examples: certain robot walkers, and analogously animals, can walk efficiently without active neural control, exploiting gravity and the structure of their body to "outsourced" aspects of locomotion [124]. In more complex animals, extended cognition includes the construction of tools, shelters, and artifacts, and in humans, this reaches the extraordinary scale of symbolic and digital technologies, from writing and mathematics to computers and the internet, which dramatically augment cognitive capacities. Beyond artifacts, humans also offload cognition onto other minds: our knowledge is largely distributed across social networks, with specialists in different fields storing and processing information that any individual could not manage alone. In all these cases, cognitive processes are not confined to the brain; they are embedded in bodies, environments, artifacts, and social structures, illustrating that scalable intelligence relies not only on internal computation but also on the clever exploitation of external physical and social structures.

Summing up, embodiment and physical constraints are fundamental to scalable intelligence in living organisms. By exploiting—rather than merely being constrained by—the laws of the physical world, organisms achieve efficient computation, reliable control, and flexible, adaptive behavior across multiple spatial and temporal scales. This principle also underlies the externalization of cognition: by offloading computations onto their bodies, environments, tools, and social networks, organisms further expand their problem-solving capabilities beyond what could be achieved internally alone.

### *Pervasive signaling enables self-organization and top-down control from goals*

A defining feature of multicellular life is the emergence of pervasive signaling: the ability of cells and tissues to communicate and coordinate across space and time. As soon as organisms become multicellular, agent-environment feedback alone is no longer sufficient for adaptive behavior. Each cell is embedded not only in its physical environment but also in the environment constituted by other cells. Consequently, behavior, decision-making, and problem-solving become fundamentally cooperative, requiring information exchange that aligns local activity with global goals. In turn, the coordinated activity of multiple, possibly heterogeneous elements vastly expands the space of problem-solving possibilities. For example, in heterogeneous neuronal networks, skewed distributions of excitability or connectivity can expand the operational space, allowing systems to explore novel solutions without compromising robustness [119]. Distributed processing across brain networks similarly enables rapid reconfiguration of large-scale neural dynamics, supporting flexible cognitive processing [125–128]. In social or swarm-like systems, agents influence each other via stigmergic cues or collective fields—such as melodic traces in a musical ensemble—enabling emergent problem-solving at scales beyond individual capacities [105].

Signaling manifests in a wide variety of channels and modalities—bioelectric, chemical, synaptic, and mechanical—but the underlying principle is consistent: communication enables distributed self-organization. Cells, tissues, and organs – agential materials at all scales [129,130] – act as competent agents, creatively interpreting incoming messages [31,114,131] while contributing to collective computation. Also molecules *and molecular dynamics*, in addition to cells, tissues, and organs, contribute to the agency of living matter. Through their context-dependent reactivity and dynamics, they contribute to the interpretation of incoming signals and to collective biological computation [97]. Biological systems leverage all the aforementioned channels and modalities to coordinate large-scale patterns of behavior, growth, and cognition, without requiring centralized control or micromanagement. Higher levels deform the option space for their parts and, metaphorically speaking, all biological subsystems are constantly "hacking" their environment, their neighbors, and their own parts using influences at all levels ranging from applying direct force to providing behavior-shaping rewards and incentives. In this sense, most intelligence in biology can be seen as fundamentally collective: this is evident from cell collectives to the most sophisticated "intelligence-enabling" biological machines: our brains, where cognition and behavior emerge from the collective organization of approximately 86 billion neurons. The importance of the collective lies not simply in the presence of interacting parts, but in their integration and alignment. Pervasive, reciprocal signaling and interactions with the environment coordinate partially autonomous components, allowing the collective to achieve capabilities that extend beyond those of its individual parts.

Crucially, the "geometry" created by pervasive signaling provides guidance for the collective – serving as a collective goal. A key design principle enabled by pervasive signaling is that it is possible to specify goals at the (higher) level of a field or ensemble, while solutions are realized by the coordinated actions of individual agents. Consider for example that a tail grafted to the flank of a salamander slowly remodels to a limb, a structure more appropriate for its new location, with tail tip cells that slowly become fingers, illustrating shape homeostasis flexibly reoriented towards a normal amphibian body plan [59,132] (Figure 7). This illustrates not only context-sensitive problem-solving, but also an important principle of top-down causation in the architecture of life. There is no local damage or other reason for tail tip cells to become remodeled into fingers: the molecular biology changes needed to make it happen are driven by highly abstract anatomical goals (pattern memories at the level of the entire body) of which the individual cells know nothing. Of course, this basic principle of making the chemistry dance to the tune of an abstract, large-scale, high-level goal is fundamental in cognitive science. A human getting out of bed in the morning to pursue her financial, social, and research goals – voluntary motion – exploits the fundamental nature of the body in which higher levels transduce their abstract goals to bend the option space for the lower levels, eventually leading to ions crossing muscle cell membranes to make behavior happen.

Bioelectricity exemplifies the centrality of pervasive signaling and top-down control from goals specified at the collective level [133]. Bioelectricity functions as both a memory medium and a developmental guide, aligning cells into coherent structures without detailed, local instructions. For instance, bioelectric prepatterns help organize complex anatomical structures—such as eyes—by specifying where competent cells should act, effectively providing cues for local problem-solving within a global framework. Bioelectricity across all tissues serves as a kind of "cognitive glue" [12], providing memory-sharing and anonymization functions [19] that help align subunits toward higher-level goals, implement high-level memory patterns that belong to the whole and to none of its parts, and underlie setpoints in high-level problems spaces (such as anatomical space) of which the parts know nothing. Other kinds of cognitive glue include stress sharing, which enable components to all work together toward high-level goals without explicit mechanisms of altruism [134].

Remarkably, this is not only true at the cells->embryo level but at the embryos->group level as well: bioelectric signals can propagate across embryos, forming a "hyper-embryo" that solves problems more efficiently than isolated cells could, illustrating how collective fields enhance adaptive capabilities [135]. To provide another example, the collective migration and self-organization of initially identical cells during morphogenesis can be guided by the field of bioelectric signals emitted by all the cells—until each cell finds its proper location within the body morphology and begins coherently emitting and sensing the signals it expects [82].

Beyond development, similar principles appear in neural circuits. The theory of "spatial computing" suggests that neural populations oscillating at slower (alpha/beta) rhythms provide flexible top-down control over incoming sensory processing, conveyed by neural populations oscillating at faster (gamma) rhythms, by selectively inhibiting spatially precise brain areas that are not processing useful information [136]. These examples illustrate that the spatial arrangement of signals—whether in bioelectric fields, chemical gradients, or neural activity—encodes constraints that define goals and regularize problem spaces. These constraints reduce the arbitrariness of computation, guide adaptive trajectories towards goals, and scaffold the emergence of coherent global behavior from local interactions. In this way, cells and neurons generate and follow manifolds: structured fields of constraints that both delimit the space of possible solutions and guide flexible problem-solving towards goals specified at the collective level.

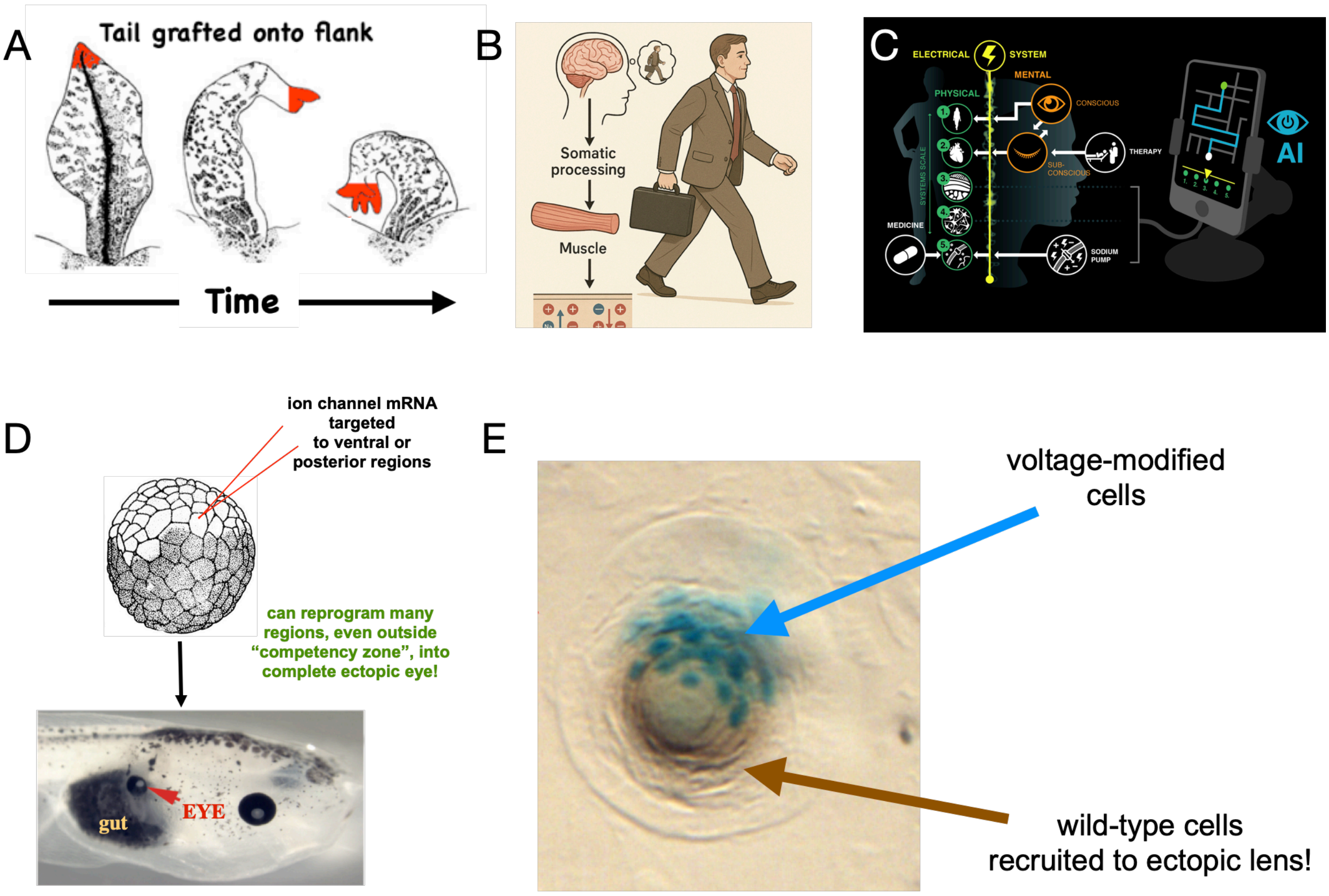

***Figure 7. Pervasive signaling enables top-down control from goals**. (A) A tail grafted to the flank of an amphibian body slowly remodels into a limb* [132]. *The most remarkable aspect is noted by focusing on the red cells – tail-tip cells sitting at the end of a tail; there is no local damage but they turn into fingers. No individual cell knows what a finger is, but the collective can tell that the current state does not match the body-level target morphology and issues commands that transduce down through the levels of organization into the molecular signals needed to convert tail tip cells into the right number of fingers. (B) This top-down control is also seen in neuroscience, where (for example) highly abstract social, financial, and other goals become transduced by the physiological networks comprising the heterarchy of the body into changes of ion flux across cell membranes needed to activate muscle cells and execute behavior to reach those goals. In other words, whether in the body or in the mind, abstract high-level conceptual structures change the course of chemistry. (C) Tools now exist to interface living tissue and artificial systems (e.g., brain computer interfaces and AI methods to decode visual images, movement intentions and speech from brain activity* [137–139]*), which means that communication can be established with many of the scales within the living hierarchy. This opens the opportunity for biomedicine and bioengineering to take advantage of the many diverse, collective intelligence modules present in vivo. (D) One example is illustrated by bioelectric control at the organ-level. Ion channel-encoding mRNA can be injected into cells (in this case, frog embryo blastomeres) to establish specific bioelectric patterns in the body. Such patterns are interpreted by surrounding cells as instructions to make organs such as ectopic eyes [116]. Importantly, this effect is not direct control of gene expression or stem cell biology – the bioengineer communicates a high-level organ goal “build an eye” (a top-level prompt) and, as befits a multiscale architecture, the lower levels implement accordingly. (E) The competency of the living material is seen when for example only a few cells are injected (shown with blue lineage label in this cross-section through an ectopic lens) – they then autonomously recruit normal (uninjected) neighbors, to complete the job as needed, by imposing a new morphogenetic goal state on the cellular collective*

*(having to first overcome the native cancer-suppression mechanisms, which cause them to resist unexpected goal states from neighbors).*

Pervasive signaling also allows biological systems to specify goals and coordinate heterogeneous, agentic components not only horizontally—i.e., across similar elements—but also across scales, which is particularly relevant given the multiscale architecture of living organisms. Evolution forces each scale to act as a "master hacker," continuously propagating coherent goals downward to subcomponents that are unaware of the larger context. Parts of organisms are constantly at risk of, metaphorically speaking, "rebelling" or pursuing local objectives, yet hierarchical organization and self-assembly ensure that the whole functions adaptively. This hierarchical coordination allows goals specified at higher levels to shape activity at lower levels without micromanaging each component, ensuring both flexibility and robustness. Local agents can act autonomously within their constraints, yet local actions are constrained and aligned by global fields, yielding coherent collective behavior, enabling adaptive behavior across scales. Moreover, the boundaries between agents (both horizontally and vertically) are plastic in living materials – they can shift over time, as a function of experiences and physiological states, continuously melting and re-forming boundaries between aligned goal-seeking subunits and their neighbors and subunits.

Pervasive signaling directly contributes to the expansion of the cognitive light cone and the nesting of Markov blankets (Figure 4). By enabling individual agents—cells, neurons, or tissues—to communicate and coordinate, signaling allows local adaptive processes to be integrated into larger-scale predictive and control structures. Each agent maintains its own "internal states" and regulatory goals, yet these are constrained and guided by collective fields, creating hierarchical layers of autonomy. Bioelectric and chemical signals, synaptic activity, and other forms of communication act as the connective tissue that links local problem-solving to organism- or tissue-level objectives, allowing the system to predict, control, and respond to events that occur across extended spatial and temporal scales. In this way, pervasive signaling transforms distributed, autonomous components into coherent, multiscale intelligence, where higher-order behaviors emerge from the interaction of nested, locally competent agents. In other words, in this perspective, intelligence is not merely a property of individual components, but emerges from the interplay of distributed signaling, physical constraints, and multiscale coordination.

One of the risks of a multiscale competency architecture is the risk of defection. A collective must constantly work to keep the option space of its parts shaped in such a way that their reward functions and sensory inputs ensure that their capacity to pursue their own agendas ultimately benefits large-scale goals of which they know nothing. In living organisms, this manifests as the ubiquitous possibility of a kind of dissociative identity disorder of the cellular collective intelligence: cancer. In this common failure mode, individual cells disconnect from the physiological network (and its grandiose organ-level setpoints) and begin to pursue ancient, single-cell goals – survival, migration, and reproduction (metastasis) [103,140]. For now, robot cancer is not an issue, because we largely make systems whose abstraction layers and error correcting codes ensure that high-level problem-solving is implemented via parts with very little autonomy, but that can change in the future.

Summing up, pervasive signaling illustrates how life exploits communication and collective coordination to achieve robustness, flexibility, and scalability. By distributing information across agents and integrating local and global processes, living organisms solve problems far beyond the capacity of individual components. Most importantly, pervasive signaling allows for the specification of collective goals—implemented, for example, in bioelectric networks or the coordinated activity of large-scale neural populations—that guide the behavior of autonomous smaller components, enabling adaptive, goal-directed activity across multiple scales and levels of organization.

## What can we learn from the five design principles?

Problem-solving in biology occurs at every level, from single cells to complex organisms. Evolution has produced systems with scalable problem-solving capabilities—a "recipe" for intelligence that relies on the progressive expansion of predictive and control capacities through five design principles. From these design principles, we can derive important insights for machine intelligence.

The *first design principle* is that living organisms are "natural born autonomous." Current machine learning, by contrast, largely sidesteps autonomy, often treating it as a feature to be added later. Indeed, engineers work very hard to make systems that rely on abstraction layers and error correction mechanisms that isolate each level from having to functionally manage the dynamic nature of its parts. Biological systems suggest that autonomy—a minimal form of "self" that persists over time, encompassing self-maintained goals, intrinsic modeling, autonomous learning, and goal setting—is fundamental, not optional. In biology, autonomy is tightly linked to basic survival needs: early nervous systems evolved to make autonomous decisions about biologically meaningful events and interact with the environment – and learn from it – to satisfy a basic repertoire of drives and needs, with more advanced cognitive competences emerging later. Incorporating these principles into AI could enable systems that are not only flexible but also capable of generating and sustaining their own goals and problem-solving objectives over time.

Achieving autonomy, however, is not something that can be easily added later on, but must be considered from the ways in which AI systems are trained. Living organisms learn autonomously, interactively, and continuously from experience. In contrast, the recent development of state-of-the-art generative AI systems has largely been based on systems that passively ingest large quantities of data—for example, linguistic data in large language models—without actively interacting with their environment. Although the training of these systems is often described as "self-supervised," substantial human supervision from outside is manifest in various ways: researchers select data, define tasks, monitor success, and even engineer or modify architectures during training. Even in reinforcement learning, rewards—the primary signals driving learning—are often carefully shaped by designers. This trend is now changing, and some aspects of autonomy—particularly those concerning autonomous learning from experience—are beginning to be integral part of the training. Important concepts from active learning, active inference, self-play, curriculum learning, and reinforcement learning from human feedback are already making these systems more interactive. New methods are also emerging to make the "harness" around AI models—the executable program that organizes model calls, tool use, and information flow—more autonomous, marking significant steps toward greater model autonomy. At the same time, advances in multimodal generative AI, increasingly endowed with world models, aim to further bridge the gap between AI systems and the physical world [2,88,141,142].

These are all important directions that remain to be fully developed but are already beginning to pay dividends in terms of increased autonomy – while simultaneously raising important ethical considerations. Just as living systems achieve scalable intelligence through autonomy, goal-directed learning, and embodied interaction, artificial systems could benefit from architectures that integrate in increasingly rich way self-directed exploration, active model-building, and hierarchical, temporally extended control—emphasizing learning strategies over fixed solutions and leveraging the environment as both a computational and regulatory resource.

The *second design principle* is that organisms' bodies and cognition grow through the self-assemblage of active components. In living systems, intelligence is not a monolithic feature imposed on inert matter; rather, it emerges from the coordinated activity of numerous agentic elements, each contributing to problem-solving at its own scale. Multicellular organisms, for example, are composed of cells that are themselves active, decision-making units, which collectively scaffold higher-order cognitive and behavioral capabilities. Developmentally, these capabilities are built incrementally: simpler structures and behaviors are reused, elaborated, and coordinated into more complex solutions (multicellular, cognitive, neural) by augmenting rather than fully replacing the underlying mechanisms.

This contrasts sharply with most current AI systems and robots, where a single, centralized "intelligent" system is imposed on passive, inert hardware. Learning and capability acquisition in these systems are typically treated as isolated, task-specific processes, rarely building incrementally on previously acquired skills or coordinating multiple semi-autonomous subcomponents. The current path to advanced AI is usually scaling up generative models with more data – but achieving continual learning as observed in biology is still challenging. While approaches such as curriculum learning or modular architectures begin to introduce some notion of staged or hierarchical learning, the richness and robustness of the multiscale, self-assembling approach in biology remains largely unexploited. Emulating this principle in artificial systems suggests a path toward scalable and robust intelligence: designing systems composed of interacting, semi-autonomous components whose coordination produces higher-order capabilities. Some aspects of this idea were suggested by early AI researchers [143,144], and have long been at the core of research in artificial life, complex systems, and open-ended evolution [33,145–150], but this pathway is less considered in the current mainstream paradigms [151,152]. By leveraging incremental growth, modular self-assembly, and collective problem-solving, artificial architectures could achieve flexibility, resilience, and emergent behaviors analogous to those observed in biological systems.

The *third design principle* is that organisms continuously rebuild themselves, achieving flexible adaptation through change. Life suggests that scalable intelligence does not emerge from freezing "optimal" solutions, but from systems that can reconstruct and refine themselves across multiple levels of organization. Future artificial agents may similarly benefit from architectures that prioritize reconstruction over preservation: models, memories, and even goals should be revisable within protected constraints, allowing agents to exploit change as a resource rather than merely tolerate it. Mechanisms for goal protection, active relearning, and multiscale self-repair could enable machines to expand their capabilities over time while maintaining coherent agency, even in fundamentally novel environments.

In contrast, most current machine learning methods struggle with changing demands, out-of-distribution tasks, and catastrophic forgetting, making them fragile under conditions of continuous change. This fragility stems in part from the assumption that each system state has a single, privileged meaning, typically imposed externally. Some current engineering solutions, such as modular architectures, curriculum learning, and continual learning, partly address these challenges, but we still lack robust methods for enabling continual reconstruction and self-production that can tolerate, or even exploit, change. Likewise, a system that can reconfigure its sensors and effectors challenges the conventional separation between hardware and software: if both the program and the hardware that implement and interpret it are continually changing, it becomes unclear how to keep them aligned, under what conditions the program remains the same or changes its function or purpose, and how to reliably evaluate or control the resulting system. By contrast, biological systems embrace the unreliable and polycomputational substrate of life [131], assigning meaning to states internally and locally. Multiple subsystems can interpret the same physical signals differently, exploiting variability and redundancy to achieve robustness and adaptability also in the faces of changes to the problem space. This capacity to assign and revise meaning flexibly underpins the resilience and scalability of living intelligence, suggesting a critical lesson for the design of future AI: to thrive in open-ended, evolving problem spaces, systems must be capable of continual reconstruction, interpretation, and self-directed adaptation.

The *fourth design principle* is that living organisms are embodied and consistently exploit physical forces and constraints, as well as mathematical and computational laws that facilitate dynamics that evolution exploits. Acting under severe, biology-like limitations—such as extreme energy efficiency, relatively slow neural processing, and "mortal computation" [153]—might seem restrictive compared with purely digital systems, which can leverage different hardware, vast energy resources, and the potential to "copy" intelligence across multiple embodiments. One could therefore argue that future AI need not be developed under the same strict constraints that shaped living organisms. However, the key point is that for biological systems, embodied and physical constraints are not merely limitations—they are a source of strength, guiding organization, efficiency, and robustness. Far from being obstacles, these constraints act as informational scaffolds, reducing arbitrariness, structuring problem spaces, and enabling scalable, adaptive intelligence. Biology offers abundant examples: traveling waves in neural tissue organize information flow, bioelectric fields guide morphogenesis, and biomechanics shape sensorimotor control.

Future AI systems could learn a similar lesson: rather than replicating the specific conditions of biological evolution, they can exploit constraints as a design resource. Just as living systems use constrained information processing—producing outputs that satisfy physical, computational, or energetic boundaries—to ensure feasibility, stability, and robustness, AI systems can leverage problem spaces structured by boundary conditions, conservation laws, or network architectures. Early examples already exist: neuromorphic hardware exploits energy-efficient architectures, and physics-informed neural networks embed physical laws to regularize predictions, much as biology uses natural laws to shape behavior and development. Yet such approaches remain marginal in mainstream AI. Fully recognizing and leveraging constraints—energetic, chemical, mechanical, and computational—offers a powerful blueprint for designing artificial systems that are robust, flexible, and capable of scalable, adaptive problem-solving. This approach becomes even more compelling in hybrid natural–artificial intelligence, where biological systems (e.g., brain–computer interfaces or cultured neural networks) are integrated with digital AI to exploit both the embodied, adaptive intelligence of biology and the new capabilities of digital systems.

The *fifth design principle* is that signaling is pervasive, and the organization of signals matters: goals specified at the collective level can exert top-down control over local components. In living systems, communication occurs across scales and modalities—from chemical gradients in single cells to bioelectric fields guiding morphogenesis, and from synaptic signaling in neural circuits to coordinated activity across entire brains. Signaling allows heterogeneous, agentic components to coordinate their actions, align local behaviors with global objectives, and adapt to changing circumstances. Crucially, goals can be specified at higher organizational levels, while the realization of solutions is distributed across lower-level components, each acting autonomously yet constrained and guided by the collective context.

In contrast, most current AI systems, including artificial neural networks, rely on narrow, predefined architectures and fixed signaling rules. Connections between nodes, learning rules, and update dynamics are typically hard-coded or globally uniform, with limited scope for autonomous adaptation at the component level. These systems generally lack the capacity for hierarchical top-down control that can dynamically shape lower-level processing in response to global goals, as observed in biological neuronal architectures. Biological systems, by contrast, leverage signaling to create structured fields, feedback loops, and collective manifolds, enabling robustness, scalability, and emergent problem-solving. Examples include bioelectric prepatterns coordinating morphogenesis, slow oscillations in neural populations imposing top-down constraints on faster rhythms, and chemical gradients guiding collective cell migration—all illustrating how pervasive signaling aligns distributed agents toward shared objectives. Recognizing the centrality of signaling in biological intelligence suggests that future artificial systems could benefit from architectures in which local agents retain autonomy but are dynamically constrained and coordinated by higher-level goals, allowing flexible, adaptive, and scalable problem-solving across multiple levels of organization.
Summing up, we have highlighted a selection of insights for machine intelligence derived from five design principles we have identified in biology. It is worth noting that the five principles we identified are neither comprehensive nor homogeneous: they operate at different levels of abstraction. Yet, taken together, they form complementary pieces of the same puzzle, spanning from *what* nature builds—autonomous and embodied systems that flexibly achieve goals and scale their capabilities—to *how* it does so—through self-assembly of components coordinated by pervasive signaling—and *under which operating conditions*—through continuous rebuilding and adaptation to change.

The autonomy principle provides a normative starting point, suggesting a design philosophy that begins with autonomy rather than with specific competences to which autonomy is added later. Embodiment is complementary, suggesting that intelligence should develop through interaction with the physical world, grounding knowledge and skills in experience. Self-assembly and pervasive signaling, considered together, highlight key mechanisms through which capabilities can expand: intelligence can emerge from the coordinated activity of numerous simple, agentic elements, while pervasive signaling enables their coordination across components and organizational levels, allowing goals to be specified at higher levels of organization. Finally, the principle of continuous rebuilding emphasizes adaptation through change, a key requirement for operating in open-ended, evolving problem spaces and environments. Living organisms not only adapt to such changing conditions but also contribute to them, continuously developing novel capabilities and thereby providing a distinctive mechanism for proliferating and scaling intelligence in the world they inhabit and co-create.

The heterogeneity of these design principles is therefore not a limitation but a source of richness: together, they provide complementary perspectives on the architecture, mechanisms, and operating conditions that enable intelligence to remain flexible, adaptive, and scalable – which can inspire future machine intelligence (see also Table 1 for example mappings between biological competences, existing artificial analogues, what is currently missing, and stronger life-inspired milestones).

**Table 1. Some illustrative examples mapping biological competences to their artificial analogues.**

| Biological principle / competence | Examples of existing artificial analogues | What is currently missing | Stronger life-inspired milestone |
|---|---|---|---|
| Autonomy | Active learning, active inference, self-play | Often the bulk of learning is not active; not tied to autopoietic or allostatic needs | Truly autonomous learning, maintenance of viable conditions and autopoiesis |
| Self-assemblage of active components | Modular and self-constructing robots | Fabrication and component semantics remain external; limited range of multiscale assembly | Self-fabrication, endogenous coordination and recombination of component roles |
| Reconstruction | Continual and curriculum learning | Architecture and loss function normally remain externally established | Viable reconstruction of the system's own problem space and measurement–control relations [148,149] |

**Towards an empirical research program to understand and identify intelligence in living organisms and machines**

A topic underlying this manuscript has been that "intelligence" is a slippery concept, and we envisage a research program—including theoretical, computational, and empirical studies—to understand and identify intelligence across biological organisms and machines. Addressing intelligence as an empirical question has the advantage that it allows us to move beyond debates over definitions and instead investigate the capacities that different systems actually exhibit, how these capacities vary across systems, and how they can be experimentally measured and compared.

We highlighted as a starting point William James' definition of intelligent systems, which focuses on flexible problem-solving—pursuing a fixed goal through variable means—and extended it to also address scalability, including the capacity to confront novel problems and autonomously create and pursue new goals, through open-ended curious exploration. We also highlighted the importance of autonomy—the first "principle" in our list—which in biological organisms (but not necessarily in artificial systems) is a precondition for everything else: without the capacity to self-sustain one's existence, the organism would not exist in the first place.

An empirical research program could start from these concepts to investigate the spectrum of flexibility, scalability, and autonomy across intelligent systems. One can imagine flexibility and scalability as two axes of a multidimensional reference space, with autonomy as a useful third axis (autonomy itself can also be graded, since different systems may have different capacities to internally maintain and persist in their goals, protect their primary goals and viability conditions, and acquire or sustain agency). One could then design experiments to situate candidate intelligent systems within this reference space and investigate their graded capacities along each axis, while potentially adding further dimensions—perhaps motivated by the other design principles we highlighted—as needed.

The research program should be equipped with empirically grounded methods to establish, for example, whether systems generate their own goals or inherit them from others, how flexibly they perform specific tasks, and precisely what competence is being claimed and from whose perspective—an external observer's or the system's own [131,154].

Empirical research should also highlight systems that exhibit high flexibility within a certain problem space but low scalability and autonomy, if their variables and objectives are supplied by an external designer and must be continuously supplied because the systems cannot generate or sustain them internally. Scalability would be significantly higher in intelligent systems capable of acquiring or reconstructing competencies beyond fixed response repertoires, and higher still in open-ended systems capable of altering the relevant variables, goals, constraints, or problem spaces themselves.

This flexibility–scalability distinction is particularly important when evaluating intelligent machines. Existing artificial systems, such as those based on generative AI, may exhibit substantial flexibility within externally established task spaces. A common pipeline for training them supplies, from the outside, training distributions, sensory variables, action spaces, reward functions, and interpretations. Yet this is rapidly changing, as methods that enable the autonomous scaling of capabilities—such as active learning, curiosity, self-play, and curriculum learning—are increasingly being incorporated into modern AI architectures, building on approaches that have long been considered in machine learning [33,155–158].

When considering autonomy and agency, biological organisms would plausibly score highly (whatever their flexibility and scalability) because—as noted above—they participate in producing and maintaining the conditions of their own existence, at fast timescales (e.g., through metabolism) and over longer timescales (e.g., through niche construction). Their internally maintained organization, together with increasingly sophisticated predictive mechanisms, enables embodied coupling with the environment, providing the substrate over which advanced cognitive abilities are acquired.

This dimension of autonomy, encompassing self-regulation, reactive (homeostatic) and anticipatory (allostatic) coupling with the environment, and predictive interoceptive processing in the service of the autonomous fulfillment of needs and goals, is more rarely addressed in machine intelligence research. Nevertheless, particular aspects of it—such as the ability to maintain a coherent context and coherent goals while operating over prolonged periods—are already recognized as important in advanced systems such as large reasoning models and are foundational to the definition of "AI agents." Experimental methods for measuring autonomy and agency should therefore be developed to assess these constructs across both biological and nonbiological realizations. Importantly, the central role of the observer must be acknowledged, where a privileged description is often tacitly assumed. Analysis must be done with explicit specification of perspective (choice of what to pay attention to, how to coarse-grain, and how to map physical observations to conceptual structures), inclusion of the system's own internal perspective on itself and its world, and an acknowledgement that any assessment of competencies is a 2-way test of IQ and imagination.

Summing up, we envisage an empirical research program to assess machine intelligence, going beyond observable success on benchmarks to establish more general cognitive capacities. Claims about flexibility, scalability, autonomy, and other competences in machines could be supported using methods similar to those employed to assess these capacities in biological organisms, including perturbation studies, robustness tests, adversarial controls, failure analysis, and tests of whether competences generalize across altered conditions. Related proposals emphasize the usefulness of applying methods from comparative and developmental psychology to AI evaluation [159,160]. For example, [161] proposes five relevant principles: adversarial control conditions, robustness to variations in stimuli, analysis of failure types, separation of mechanism from behavior, and evaluation of a system in ways appropriate to its particular modalities and limitations. Furthermore, [162] advances six principles, inspired by developmental and comparative psychology, for rigorous evaluation of AI systems, highlighting the need to account for anthropomorphic bias, distinguish performance from competence, establish construct validity, and recognize the value of negative results. We envisage the importance of integrating and extending these converging proposals into an integrative framework for systematically assessing flexibility, scalability, autonomy, and other dimensions of intelligence across biological and artificial systems.

Finally, such a research program would benefit from solid conceptual foundations—theoretical and formal tools that remain to be fully developed to define, measure, and quantify forms of intelligence that may manifest in diverse and unexpected ways. Here, we offered some starting points, such as the notion of a "cognitive light cone," which characterizes the extent of a system's goal-directed capacities across space, time, and other dimensions, and the notion of a Markov Blanket, which formalizes the boundary between an organism and its environment as a statistical partition between internal, sensory, active, and external variables, potentially across nested levels. We have also illustrated these concepts in the context of biological organisms. Yet these are by no means the only possible theoretical tools, and have been subject to criticism. For example, some have emphasized the need for caution when applying concepts such as Markov Blankets, and particularly nested Markov Blankets, to biological categories, and it remains debated to what extent this formalism naturally extends from the cellular to the organismal and social levels [163–166]. One appeal of such formal tools is that, like William James' notion of intelligence as flexible and scalable problem-solving and related attempts to define fluid intelligence [32,34], they offer a framework for studying diverse forms of intelligence across biological and nonbiological realms without starting from anthropomorphic assumptions. At the same time, it remains an open question to what extent these concepts can be generalized, adapted, or ultimately need to be replaced to fully capture the diversity of intelligence across natural, synthetic, and hybrid embodiments.

## Conclusion

Here, we started from the consideration that life achieves flexible and scalable intelligence. Importantly, flexibility and scalability in biology are precise concepts: flexibility implies that organisms display general competences relative to the range of problems they encounter in their environments, able to learn many new strategies and respond to novel challenges, yet constrained by the boundaries of their niche. Scalability implies that, under certain conditions, living organisms can autonomously invent new goals, expand their problem-solving capacities, and enlarge their "cognitive light cone," with corresponding predictive and control abilities over larger portions of space and time.

We then reviewed five key design principles that illustrate how biological systems achieve flexible and scalable intelligence: autonomy, multiscale self-assembly, continuous rebuilding, exploitation of physical constraints, and pervasive signaling. These principles show that intelligence in living organisms is not about having pre-specified solutions to every problem, but about building systems capable of learning, adapting, and coordinating across levels of organization within their ecological and physical niches, while being guided by goals. We also considered ways in which future AI systems could take inspiration from these principles to design more advanced systems, and how adopting theoretical and empirical methods to study biological intelligence could also help us probe, identify, and understand machine intelligence.

Translating these lessons to machine intelligence suggests that the objective is not to replicate every detail of biological intelligence, but to emulate its ability to flexibly achieve goals, scale up problem-solving capacities, coordinate autonomous components, and adapt robustly across scales—achieving resilience and intelligence that grows with experience and context. In essence, the challenge for artificial intelligence is not to copy life, but perhaps to continue it.

## Acknowledgements

This research received funding from the European Research Council under the Grant Agreement No. 820213 (ThinkAhead); the Italian National Recovery and Resilience Plan (NRRP), M4C2, funded by the European Union, NextGenerationEU (Project IR0000011, CUP B51E22000150006, "EBRAINSItaly"; Project PE0000013, CUP B53C22003630006, "FAIR"; Project PE0000013, CUP J93C24000320007, "FAIR"; Project PE0000006, CUP J33C22002970002 "MNESYS"), the Ministry of University and Research, PRIN PNRR P20224FESY, PRIN 20229Z7M8N, PRIN 2022EBC78W, and Eugene Jhong. This publication was also made possible through the support of Grant 62212 from the John Templeton Foundation. The opinions expressed in this publication are those of the author(s) and do not necessarily reflect the views of the John Templeton Foundation. The funders had no role in study design, data collection and analysis, decision to publish, or preparation of the manuscript. We used a Generative AI model to correct typographical errors and edit language for clarity. We are grateful to the anonymous reviewers for their valuable feedback and constructive suggestions.